\documentclass[aps,prb,reprint,superscriptaddress,letter,floatfix,preprintnumbers,longbibliography]{revtex4-1}

\usepackage[T1]{fontenc}

\usepackage{graphicx}

\usepackage[version=3]{mhchem}

\usepackage{units}

\usepackage[charter,greekuppercase=italicized]{mathdesign}

\newcommand{\muB}{\mu_\text{B}}
\newcommand{\muO}{\mu_0}

\newcommand{\TN}{T_\text{N}}

\newcommand{\TSDW}{T_\text{SDW}}
\newcommand{\Thp}{\theta_\text{p}}

\newcommand{\Hcr}{H_\text{cr}}

\newcommand{\musat}{\mu_\text{sat}}
\newcommand{\meff}{\mu_\text{eff}}

\newcommand{\Hparc}{H\parallel c}
\newcommand{\Hperpc}{H\perp c}

\newcommand{\Hac}{H_\text{ac}}

\newcommand{\chiO}{\chi_0}
\newcommand{\dd}{\mathrm{d}}

\newcommand{\EuCa}{\ce{Eu_{0.88}\-Ca_{0.12}\-Fe2\-As2}}
\newcommand{\EuCaCa}{\ce{Eu_{0.57}\-Ca_{0.43}\-Fe2\-As2}}
\newcommand{\EuCaFeAs}{\ce{Eu_{1-$x$}\-Ca_$x$\-Fe2\-As2}}
\newcommand{\Eu}{\ce{EuFe2As2}}
\newcommand{\Ca}{\ce{CaFe2As2}}
\newcommand{\AFeAs}{\ensuremath{A}\ce{Fe2As2}}

\newcommand{\cax}{c\text{-axis}}
\newcommand{\abplane}{ab\text{-plane}}

\newcommand{\oC}{^{\circ}\text{C}}

\newcommand{\Euion}{\ce{Eu^{2+}}}
\newcommand{\Feion}{\ce{Fe^{2+}}}
\newcommand{\Caion}{\ce{Ca^{2+}}}

\usepackage[colorlinks,plainpages=false,linkcolor=blue,urlcolor=blue,citecolor=blue,pdfpagemode=UseNone,pdfstartview=FitBH]{hyperref}

\begin{document}

\preprint{Ver. \today}

\title{Magnetic phase diagram of Ca-substituted \Eu{}}

\author{L.\ M.\ Tran}
\email[Corresponding author: \vspace{8pt}]{l.m.tran@int.pan.wroc.pl}
\affiliation{Institute of Low Temperature and Structure Research, Polish Academy of Sciences, P.O. Box 1410, PL-50-422 Wroclaw, Poland}

\author{M.\ Babij}
\affiliation{Institute of Low Temperature and Structure Research, Polish Academy of Sciences, P.O. Box 1410, PL-50-422 Wroclaw, Poland}

\author{L.\ Korosec}
\affiliation{Laboratorium f\"ur Festk\"orperphysik, ETH Z\"urich, CH-8093 Zurich, Switzerland}

\author{T.\ Shang}
\affiliation{Paul Scherrer Institut, CH-5232 Villigen PSI, Switzerland}
\affiliation{Institute of Condensed Matter Physics, \'Ecole Polytechnique F\'ed\'erale de Lausanne (EPFL), Lausanne CH-1015, Switzerland}

\author{Z.\ Bukowski}
\affiliation{Institute of Low Temperature and Structure Research, Polish Academy of Sciences, P.O. Box 1410, PL-50-422 Wroclaw, Poland}

\author{T.\ Shiroka}
\affiliation{Laboratorium f\"ur Festk\"orperphysik, ETH Z\"urich, CH-8093 Zurich, Switzerland}
\affiliation{Paul Scherrer Institut, CH-5232 Villigen PSI, Switzerland}

\date{\today}

\begin{abstract}
The simultaneous presence of a Fe-related spin-density wave and antiferromagnetic order of
Eu$^{2+}$ moments ranks EuFe$_2$As$_2$ among the most interesting parent compounds of iron-based pnictide superconductors. Here we explore the consequences of the dilution of Eu$^{2+}$ magnetic lattice through on-site Ca substitution.
By employing macro- and microscopic techniques, including electrical transport and magnetometry, as well as muon-spin spectroscopy, we study the
evolution of Eu magnetic order in both the weak and strong
dilution regimes, achieved for Ca concentration $x(\mathrm{Ca}) = 0.12$ and 0.43, respectively.
We demonstrate the localized character of the Eu antiferromagnetism
mediated via RKKY interactions, in contrast with the largely
itinerant nature of Fe magnetic interactions.
Our results suggest a weak coupling between the Fe and Eu magnetic sublattices and a rapid decrease of the Eu magnetic interaction strength upon Ca substitution. The latter is confirmed both by the depression of
the ordering temperature of the \Euion{} moments, $T_\mathrm{N}$,
and the decrease of magnetic volume fraction with increasing
$x(\mathrm{Ca})$. We establish that, similarly to the \Eu{} parent compound, the investigated Ca-doped compounds have a twinned structure and undergo a permanent detwinning upon applying an external magnetic field.

\end{abstract}

\maketitle

\section{Introduction}

Superconductivity and magnetic order are among the most studied topics in contemporary solid-state physics. Although commonly considered to be antagonistic phenomena, in certain cases superconductivity has been observed to coexist
with magnetic order.\cite{Maple1995}

In this respect, iron-based superconductors (Fe-SC) are among the most recent examples to exhibit such an
interplay between superconductivity and magnetism.\cite{Vorontsov2009,Sanna2011}
In particular, \Eu-based compounds 
possess some unique properties, reflecting the simultaneous occurrence of superconductivity and of
two separate magnetically-ordered subsystems (due to iron and europium layers).\cite{Zapf2017,Tran2012NJP}
In the \Eu{} parent compound, a spin-density-wave (SDW) order associated with the Fe 3$d$-electrons is observed at
$\TSDW\approx\unit[190]{K}$. The SDW transition is accompanied by a structural phase transition from the tetragonal ($I4/mmm$) to the orthorhombic ($Fmmm$) structure.
Additionally, europium magnetic moments order antiferromagnetically at $\TN=\unit[19]{K}$
in an A-type magnetic structure.
This corresponds to magnetic moments being aligned ferromagnetically along the $a$-axis, and
antiferromagnetically along the $c$-axis.\cite{Xiao2009}
Superconductivity in the \Eu-based compounds emerges when the magnetic order of the itinerant Fe 3$d$-electrons gets suppressed, 
achieved either via applied pressure\cite{Terashima2009JPSJ-p,Miclea2009PRB-p,Morozova2015}
or chemical substitution.\cite{Zapf2017} The latter involves either isovalent substitutions in the FeAs-layers\cite{Jiang2009PRB,Tran2012NJP,Tran2012PRB,
Ren2009PRL,Jiao2013,Paramanik2013} or electron/hole doping in the Eu-layers.\cite{Jeevan2008Kdoped,Qi2008,Zhang2012}

Upon substitution in the FeAs-layers, the SDW ordering temperature decreases compared to the $\TSDW$ of the parent compound,
with SDW order being completely suppressed in the overdoped compounds. For Co-doped compounds it was reported that chemical substitution modifies the shape of the SDW,\cite{Blachowski2011Codoped} yet the doping does not change the magnetic ground state of the \Feion{} magnetic moments, which remains antiferromagnetic as long as the order is present.\cite{Jin2016}
The magnetic ground state of the Eu-subsystem, however, changes from an A-type antiferromagnet, via a canted antiferromagnet, to a ferromagnet
upon doping the FeAs-sublattice.\cite{Jin2016,Anand2015,Nowik2011P-doped}

A rather different behavior is observed for isovalent substitution of europium. On one hand, the
Ca-doping does not lead to superconductivity under ambient pressure
since the Fe-related SDW remains nearly unchanged, i.e., the substitution does not modify significantly
neither the shape of SDW, nor its ordering temperature.\cite{Komedera2018,Mitsuda2011JPSJ,Mitsuda2011,Anupam2011}
On the other hand, the dilution of the Eu-sublattice with nonmagnetic \Caion{} ions,
decreases the temperature of the antiferromagnetic order,  eventually leading to its disappearance
for Eu dilutions above 50\%.\cite{Mitsuda2011,Mitsuda2011JPSJ}
Thus, we expect 
the interactions between \Euion{} ions to change significantly 
upon Ca doping,
with \EuCaFeAs{} representing an ideal system for studying the
Eu-related magnetism in the \Eu-based compounds 
and its possible interplay with Fe-magnetism, while keepeing the SDW order unchanged.

To shed light on the magnetic phase transitions of such a system, we investigated the
zero-field resistivity, ac susceptibility, dc magnetization, and
muon-spin relaxation ($\mu$SR) 
of two Ca-doped EuFe$_2$As$_2$ compounds.
It is worth noting that the $\mu$SR spectroscopy has been successfully used to investigate other 
\Eu{}-based
compounds, for which it could reveal the nature of magnetic order, the magnetic phase
diagram, and the internal field values.\cite{Guguchia2013,Goltz2014,Anand2015}
Unlike previous studies, where only the effects of Fe magnetic lattice dilution
or As isoelectronic replacement were considered, here we investigate
the microscopic consequences of the Eu magnetic lattice dilution
via on-site substitution with nonmagnetic \Caion{} ions.
To date no microscopic studies of the Ca-doped EuFe$_2$As$_2$ system exist.
By using $\mu$SR techniques we aimed to explore the interplay of  
Fe-3$d$ and Eu-4$f$ magnetism, as well as a possible coupling between the two, with a particular
focus on the low-temperature antiferromagnetic region, dominated by the
rare-earth magnetic interactions.

\section{Experimental details\label{sec:details}}
Single crystals of \EuCaFeAs{}
were grown using the Sn\nobreakdash-flux method. The Eu, Fe, As, Ca, and Sn elements in molar ratios of $(1-x)$:2:2:$x$:30 were loaded into alumina crucibles and sealed in quartz ampules under vacuum.
To dissolve all the ingredients, the ampules were heated slowly to 1050$\oC$, kept at this temperature for several hours,
and then cooled down slowly to 650$\oC$ with a rate of $\unit[2]{\oC/h}$. Next, the liquid tin was decanted from the crucibles. The Sn residues on the crystals were removed via etching in diluted hydrochloric acid.

The chemical composition of the obtained crystals was determined using
the energy-dispersive x-ray spectroscopy (EDX), %
whereas the crystal structure and phase purity were characterized by powder x-ray diffraction (XRD)
using an X'Pert Pro powder diffractometer (PANalytical, The Netherlands) equipped with a linear PIXcel detector.

The ac magnetic susceptibility measurements were performed using an Oxford Instruments susceptometer in the
temperature range of 2--$\unit[300]{K}$ and external fields up to $\unit[2]{T}$, probing with a driving field
$\mu_0\Hac \simeq \unit[1]{mT}$ at a frequency $f=\unit[1.111]{kHz}$.
The dc magnetization measurements were carried out in a Physical Properties Measurement System (PPMS, Quantum Design) in magnetic fields up to $\unit[9]{T}$ and covering a temperature range of 2--$\unit[320]{K}$.
The magnetization data were collected in both zero-field-cooling (ZFC) and
field-cooling (FC) modes, whereas the ac magnetic susceptibility measurements were performed in ZFC mode only.
Both dc and ac measurements were conducted with the external fields applied parallel as well as perpendicular to the $\cax$.

Resistivity data were collected using the PPMS platform in the temperature range from 2 to $\unit[300]{K}$. Silver-wire contacts were mounted on the  surface parallel to the crystallographic $ab$-plane using DuPont conductive silver paint; the contact resistances were less than $\unit[0.5]{\Omega}$.
The $\mu$SR measurements were carried out at the General Purpose Spectrometer (GPS) at the $\pi$M3 beam line
of the Swiss Muon Source 
at the Paul Scherrer Institut (PSI) in Villigen, Switzerland.
The temperature dependence of the muon-spin relaxation 
was investigated both in zero-(ZF) and in longitudinally (LF) applied
magnetic fields.
Several single crystals, 0.5 to 1\,mm thick
were arranged in a mosaic configuration with their $c$-axes collinear
on top of an aluminated mylar sheet. The
latter was then folded to form an envelope on top of which suitable 50-$\mu$m thick
kapton foils were added to optimize the muon stopping rate. The sample was
finally mounted on a copper fork (fly-past setup) with a negligible
background count. Active-field compensation during ZF-$\mu$SR measurements
ensured a stray field value below 1\,$\mu$T. Data in both ZF and LF experiments
were collected in a temperature range of 1.6 to 60\,K and, occasionally, up to 240\,K
(to confirm the known Fe spin-density wave at ca.\ 190\,K). For the LF experiments
magnetic fields up to 0.6\,T were applied.
\section{Experimental results and discussion\label{sec:results}}
\subsection{Composition and crystal structure}
\begin{figure}[ht]
\includegraphics[width=0.48\textwidth]{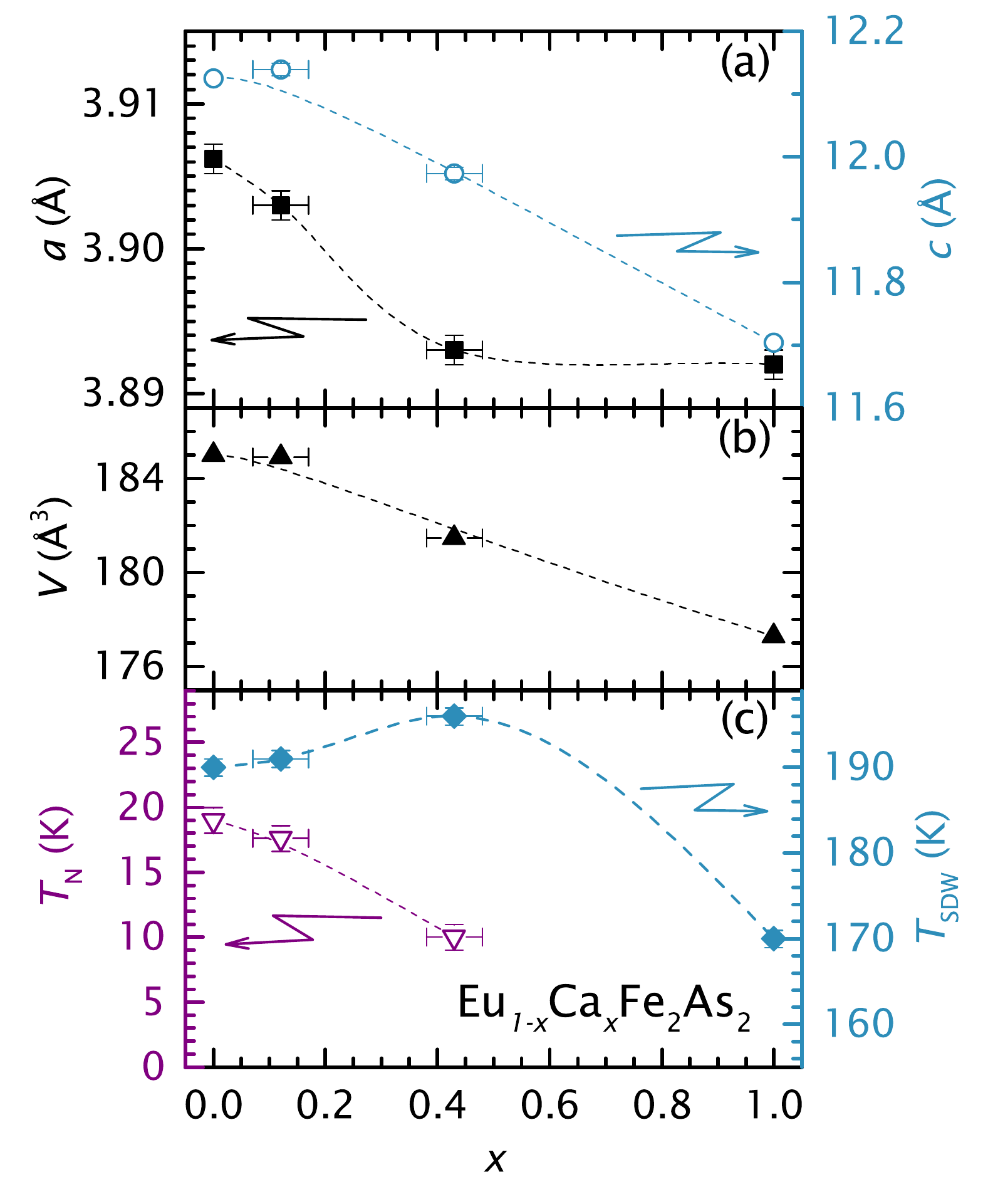}
\caption{Dependence of (a) $a$ and $c$ lattice parameters, (b) unit cell volume $V$, and (c) $\TN$ and $\TSDW$ temperatures on the Ca-concentration $x$ in \EuCaFeAs{}. Dashed lines are guides to the eye.\label{fig:Ca-x}%
}
\end{figure}

For the current study we prepared single crystals of \Eu{} and \ce{CaFe2As2}, as well as of \EuCa{} and \EuCaCa{}.
All the observed x-ray diffraction reflections for the investigated samples could be indexed to the tetragonal $\ce{ThCr2Si2}$\nobreakdash-type structure ($I4/mmm$ space group), expected for the $\AFeAs$\nobreakdash-based systems.
The refined $a$ and $c$ lattice parameters and the calculated unit cell volumes $V$ are shown in Fig.~\ref{fig:Ca-x}(a,b). Both lattice parameters (and the unit cell volume) decrease with increasing Ca-concentration and are in good agreement with those reported by Mitsuda et al.\cite{Mitsuda2011}

\subsection{Electrical resistivity\label{ssec:Res}}

\begin{figure}[t]
\includegraphics[width=0.48\textwidth]{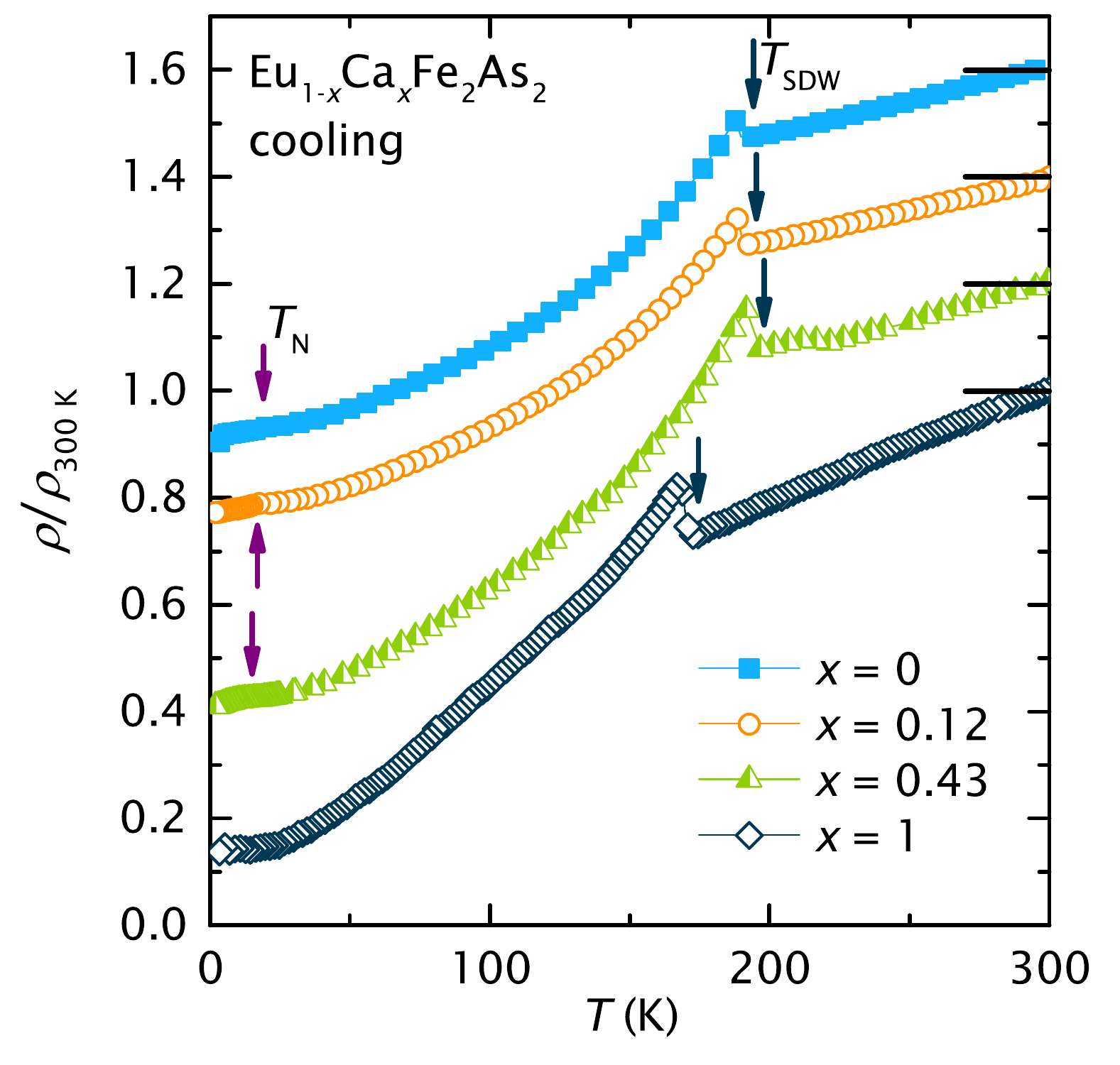}
\caption{Normalized resistivity $\rho/\rho_{300\text{K}}$ vs.\ temperature $T$ for the \EuCaFeAs{} series with $x$ = 0, 0.12, 0.43, and~1. The $\TN$ and $\TSDW$ temperatures are marked with arrows. For clarity, the resistivity curves for $x>0$ are shifted vertically by 0.2 units, with the horizontal lines marking the respective $\rho/\rho_{\unit[300]{K}} =1$ levels.\label{fig:Res}%
}
\end{figure}

The temperature dependencies of resistivity normalized to the 300-K resistivity values for the whole $\EuCaFeAs$ series ($x$ = 0, 0.12, 0.43 and 1) are shown in Fig.~\ref{fig:Res}. For all the compositions a pronounced anomaly is observed at the spin-density-wave (SDW) ordering temperature of the Fe conduction electrons~($\TSDW$), coinciding with a tetragonal-to-orthorhombic structural phase transition.
The $\TSDW\unit[\sim190]{K}$ is nearly the same for all compounds, except
for $\Ca$ ($\TSDW\sim\unit[170]{K}$), 
in good agreement with the literature data.\cite{Mitsuda2010JPSJL-p,Mitsuda2011JPSJ,Mitsuda2011}

A much weaker anomaly, corresponding to the antiferromagnetic ordering of $\Euion$ magnetic moments, is observed below \unit[19]{K}. As expected, the  dilution of the Eu magnetic sublattice (via nonmagnetic Ca ions) leads
to a decreased N\'{e}el temperature $\TN$.
The observed $\TN$ for \EuCaFeAs{} are in good agreement with those reported
by Mitsuda et al.\cite{Mitsuda2011}
%
The evolution of both the $\TSDW$ and $\TN$ values with Ca-concentration
is summarized in Fig.~\ref{fig:Ca-x}(c).

\subsection{Magnetization\label{ssec:Mag}}
\begin{figure}[h]
\includegraphics[width=0.48\textwidth]{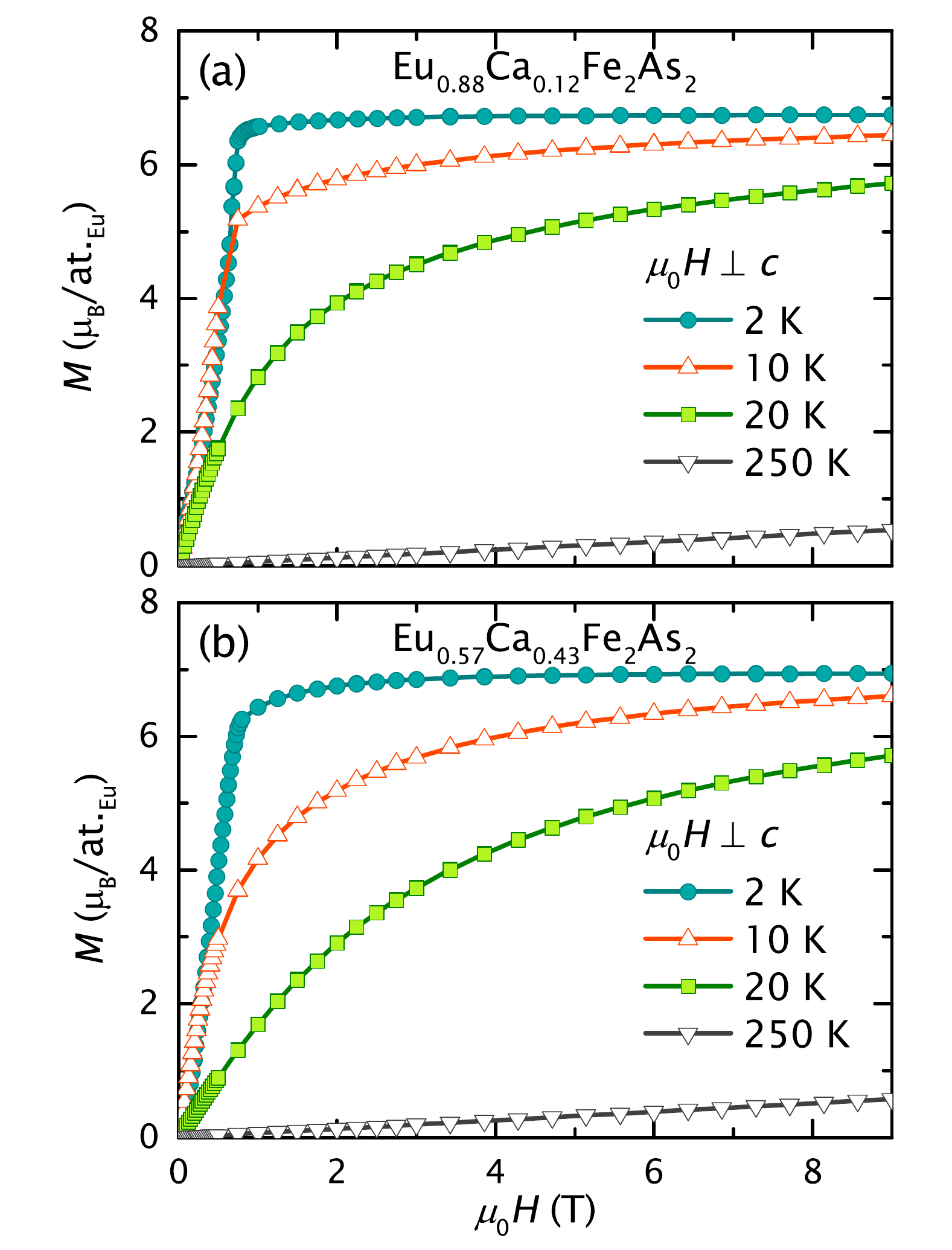}
\caption{Magnetic field dependence of magnetization per Eu atom for
(a)~\EuCa{} and (b)~\EuCaCa{} measured at selected temperatures
and in
increasing fields applied perpendicular to the $\cax$. \label{fig:M(H)}}
\end{figure}

\begin{figure}[h]
\includegraphics[width=0.48\textwidth]{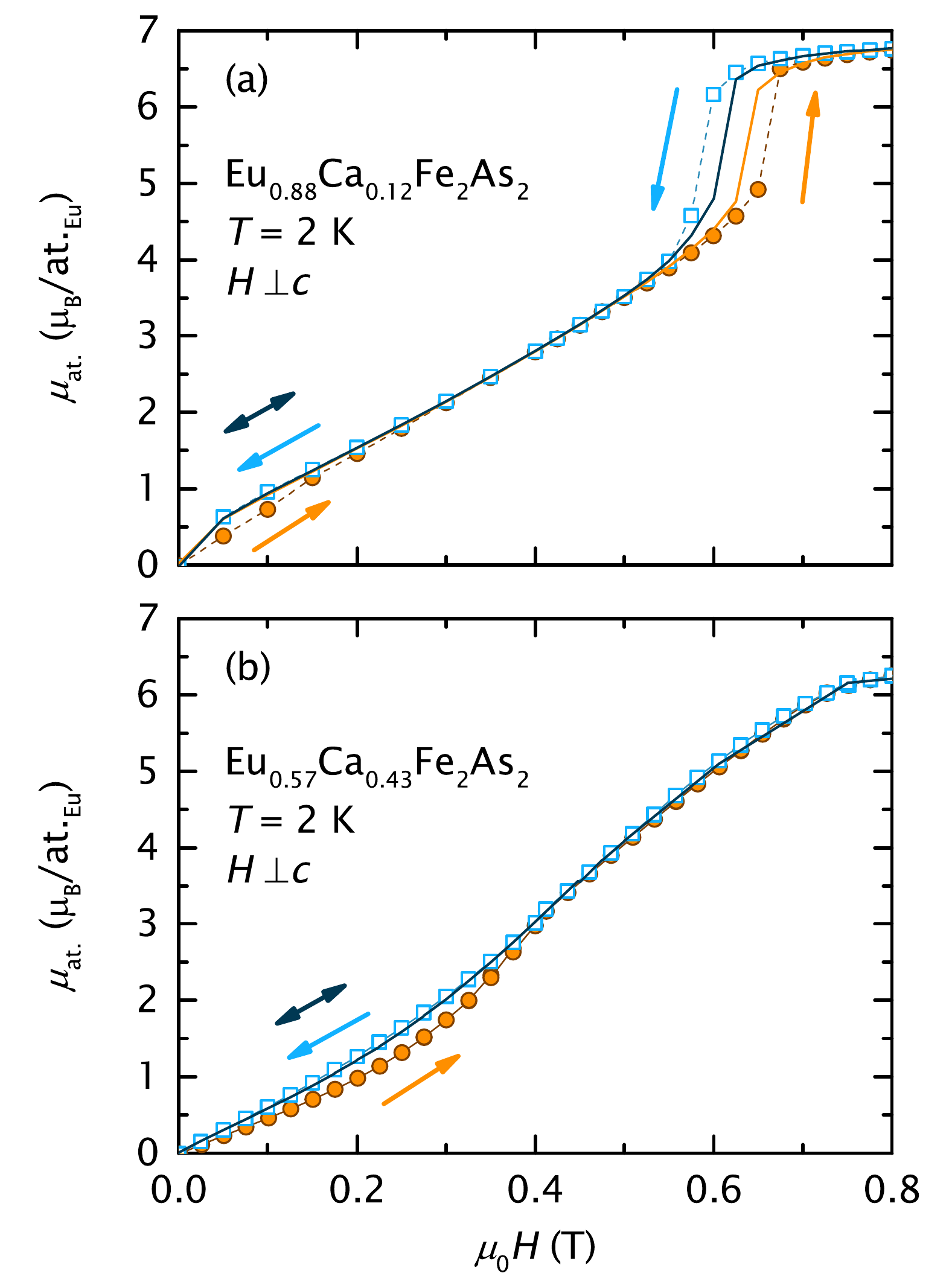}
\caption{Field dependence of 
(a) \EuCa{} and (b) \EuCaCa{} magnetization (in units of magnetic moment per Eu atom)  
measured at $\unit[2]{K}$ in increasing (full circles) and decreasing (open squares) external magnetic fields applied perpendicular to the $\cax$.
Thick solid lines represent the magnetization measured in the second run (see text) in increasing/decreasing fields.\label{fig:M(H):hyst}%
}
\end{figure}

The field-dependent magnetization $M(H)$ of both 12\% and 43\% Ca-doped compounds was investigated in external magnetic fields applied either parallel (not shown) or perpendicular (Fig.~\ref{fig:M(H)} and \ref{fig:M(H):hyst}) to the $\cax$. The measurements were carried out in both increasing and decreasing 
fields. No spontaneous magnetization was detected in any of the
compounds and the initial slope is linear, as 
expected for antiferromagnetic systems. Magnetic hysteresis was observed for both systems, however only for fields applied perpendicular to the $\cax$ and at temperatures below \unit[5]{K}.

The $M(H)$-curves measured in the magnetic fields  applied perpendicular to the $\cax$ are presented in Fig.~\ref{fig:M(H)}.
At $T=\unit[2]{K}$, the magnetization  of both \EuCa{} and \EuCaCa{} is characterized by a linear initial slope, a small magnetic hysteresis at \unit[0.2]{T}, and a metamagnetic transition at 0.6 and \unit[0.7]{T}, respectively (Fig.~\ref{fig:M(H):hyst}).
In the \EuCa{} case, an additional hysteresis is observed at $\unit[\sim 0.6]{T}$  [see Fig.~\ref{fig:M(H):hyst}(a)].

As can be seen in  Fig.~\ref{fig:M(H):hyst}(a), the field-dependent magnetization of \EuCa{} changes from linear (first field increase) to one  with a positive curvature (subsequent measurements), suggesting the presence of ferromagnetic correlations.
On the other hand, a linear dependence up to \unit[0.1]{T} is observed in \EuCaCa{} for both increase- and decrease-field measurements [Fig.~\ref{fig:M(H):hyst}(b)].

For both investigated compounds the low-field hysteresis (at $\sim$\unit[0.2]{T}) disappears in subsequently repeated increase- and decrease-field  measurements and the $M(H)$-curves follow the same dependence as the first decrease-fields dependence (cf.\ solid lines in Fig.~\ref{fig:M(H):hyst}).  Additionally, in the \EuCaCa{} case, the width of the high-field hysteresis loop (at $\sim$\unit[0.6]{T}) decreases in the subsequent measurements.
Moreover, the initial field dependence of magnetization can only be reproduced after heating the samples above the respective $\TSDW$.

A similar field dependent magnetization was found in the \Eu{} parent compound. As shown by Jiang et~al.,\cite{Jiang2009NJP} when an external magnetic field is applied perpendicular to the $\cax$, subsequent metamagnetic transitions are observed. Neutron-spectroscopy studies have revealed that \Eu{} crystals have a twinned structure, whereby the application of an external magnetic field modifies the
twinning fraction, at fields corresponding to the metamagnetic transitions.\cite{Xiao2009}  
A study by Zapf et al.\cite{Zapf2014} suggests that both the metamagnetic transitions and the hysteretic behavior are due to ``detwinning''.
There it was shown not only that \Eu{} undergoes two ``detwinning'' processes (at 0.1 and at \unit[1]{T}),
but also that the ``detwinning'' is permanent as long as the sample is not heated above $\TSDW$. A recent theoretical study indicates that the ``detwinning'' in magnetic fields is characteristic of Fe-SC systems.\cite{Maiwald2018}
However, in \Eu-based systems, the fields allowing for a persistent detwinning are much smaller 
than those required in other $A$\ce{Fe2As2}-systems ($A$ = Ba, Ca, ...), thus reflecting 
the large-spin Eu ions and the coupling of the Fe and Eu sublattices.

The close resemblance of the \EuCa{} magnetization curve to that of the undoped \Eu{} suggests that the 12\% Ca-doped compound undergoes a similar ``detwinning'' transitions as the \Eu{} parent compound.
By contrast, the magnetization of the \EuCaCa{} compound exhibits only the low-temperature magnetic hysteresis. Additionally, its $M(H)$-dependence exhibits a negative curvature in the rebound field, contrary to the \EuCa{} positive curvature.
Remarkably, such intermediate metamagnetic states were not observed in the magnetization curves of systems with substituted FeAs-sublattice that possess the canted-AF structure of  \Euion magnetic moments.\cite{Tran2012NJP,Blachowski2011Codoped,Jiang2009PRB}

For each magnetization isotherm measured at $T<\TN$, the saturation magnetization can be determined.
For instance, at $\unit[2]{K}$ the saturation magnetization $\musat$ corresponds to $\unit[7]{\muB}$, which is the expected theoretical value for magnetic
\Euion{} ions ($\musat^\text{th}=gJ$, with $g=2$ and $J=7/2$).
It was shown that in \Eu-based systems the saturation state is equivalent
to a field-induced ferromagnetic state.\cite{Jiang2009NJP,Guguchia2011,Tran2012NJP,Tran2012PRB} The field at which the system ``switches'' from an antiferromagnetic (AF) to a field-induced ferromagnetic state (FI-FM) is represented by the ``crossover'' field $\Hcr$. Such $\Hcr$ values can,
therefore, be calculated by determining the minimum in the second derivative of magnetization
$\dd^{2} M/\dd H^{2}$.


\begin{figure}[ht]
\includegraphics[width=0.48\textwidth]{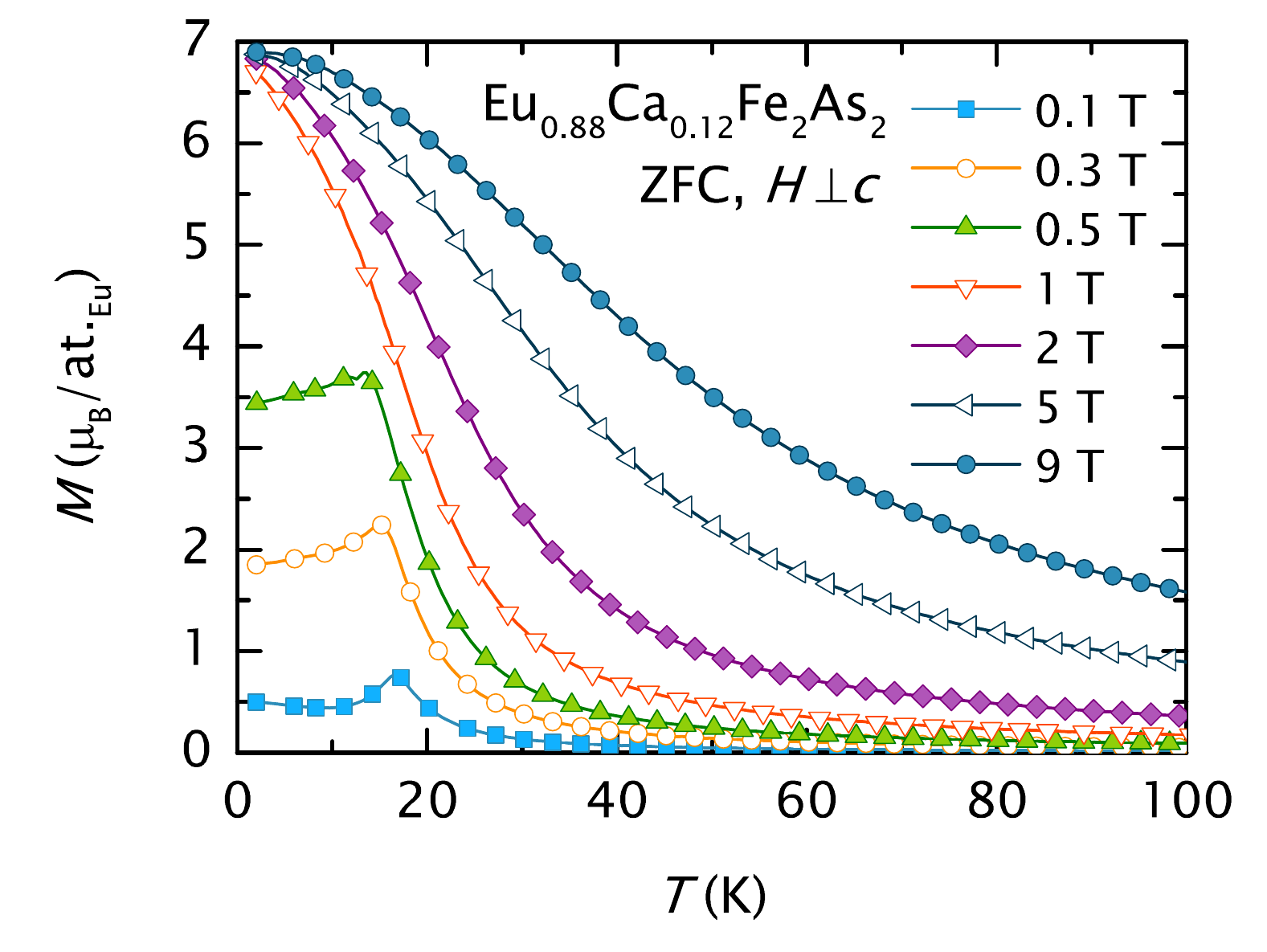}
\caption{Temperature dependence of magnetization of $\EuCa$ measured in ZFC-mode in several magnetic fields ($\muO H$) applied perpendicular to the $\cax$.\label{fig:M(T)}%
}
\end{figure}

Figure~\ref{fig:M(T)} shows the temperature dependence of $\EuCa$ ZFC
magnetization measured at some representative fields applied perpendicular to the $\cax$.
As expected for an AF transition, the maximum in $M(T)$, corresponding to
the ordering temperature of $\Euion$ magnetic moments (e.g., at	$\TN=\unit[15.6]{K}$ in $\unit[0.1]{T}$), shifts towards lower temperatures with increasing magnetic field. Hence $\TN$ was determined by using the Fisher's method --- i.e., from the maximum of the derivative of the $T\chi(T)$ product.\citep{Fisher1962}
Note, however, that for fields above $\unit[0.5]{T}$, $M(T)$ does not show a maximum, but only a smooth increase with decreasing temperature (up to $M\sim\unit[7]{\muB/Eu}$ close to $T=\unit[0]{K}$).

\begin{figure}[ht]
\includegraphics[width=0.48\textwidth]{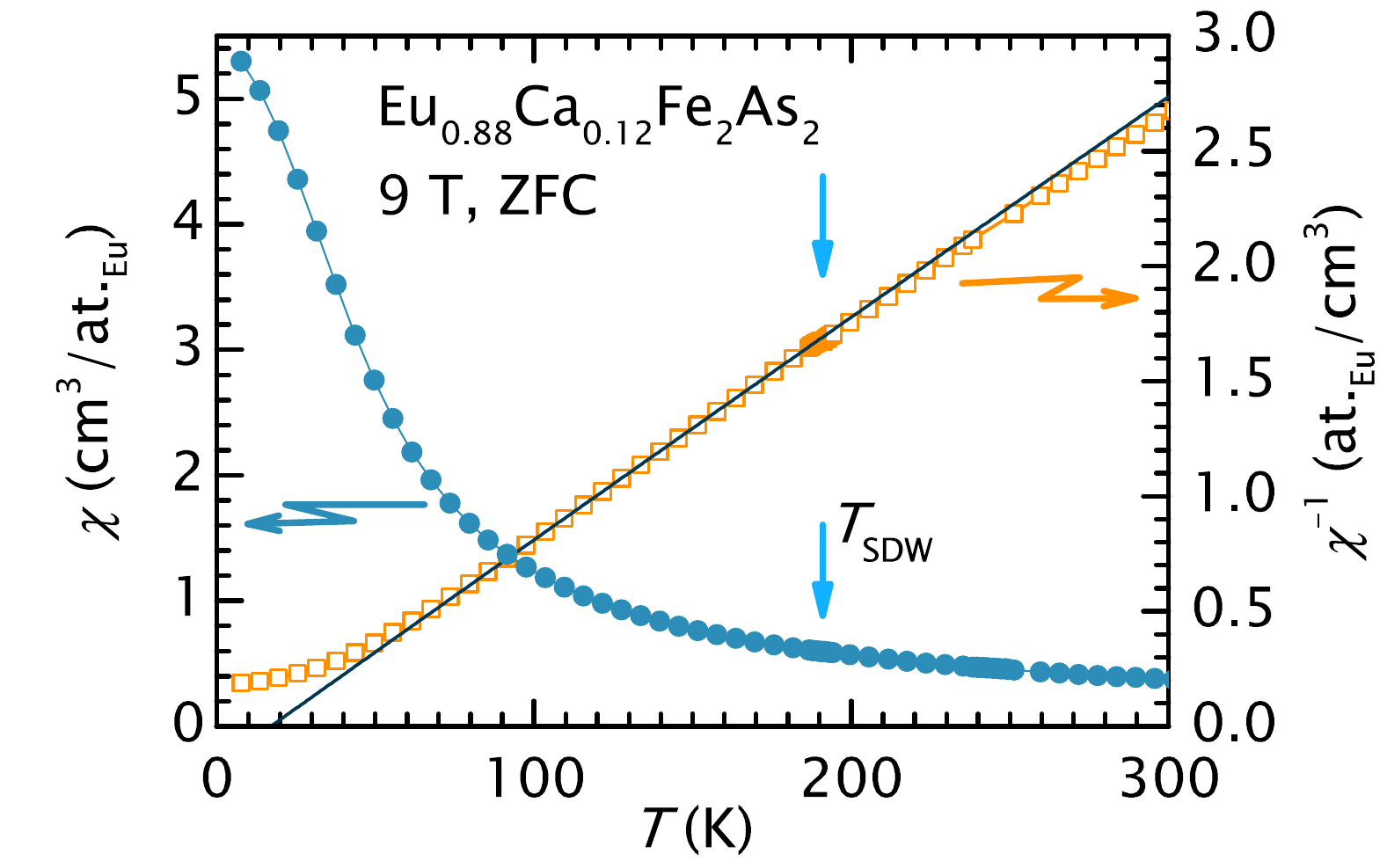}
\caption{%
Temperature-dependent ZFC magnetic susceptibility $\chi(T)$ (full circles, left axis) and its inverse $1/\chi$ (open squares, right axis) for $\EuCa$ measured in a 9-T magnetic field applied parallel to the $\cax$.
The solid line represent the fit by means of a modified Curie-Weiss law [Eq.~(\ref{eq:MCW})]
in the $100$--$\unit[180]{K}$ temperature range. The $\TSDW$ is marked with an arrow.\label{fig:Chi(T)}%
}%
\end{figure}

In contrast to the Eu-related AF transition, the anomaly associated with the Fe-SDW transition is barely visible in the magnetization measurements. And when so, it can be observed only at high  magnetic fields ($\muO H>\unit[2]{T}$), as marked with blue vertical arrows in Fig.~\ref{fig:Chi(T)}. These results are different from those of resistivity, where the SDW transition is much more prominent than the AF transition.

The 9-T dc magnetization data were used to calculate the magnetic susceptibility $\chi=M/H$. A typical temperature dependence of susceptibility and of its inverse for $\EuCa$ 
in $H \parallel c$ are shown in Fig.~\ref{fig:Chi(T)}.
In the $100$--$\unit[180]{K}$ temperature range, the $\chi(T)$ susceptibility curves of $\EuCa$ and $\EuCaCa$, measured in magnetic fields parallel and perpendicular to the $\cax$, could be fitted by means of a modified Curie-Weiss law:
\begin{equation}
\chi=\dfrac{N_{\text{A}}}{3k_{\text{B}}}\dfrac{\meff^{2}}{T-\Thp}+\chiO.\label{eq:MCW}
\end{equation}
Here $N_{\text{A}}$ is the Avogadro's constant, $k_{\text{B}}$ the Boltzmann constant,
$\Thp$ the Weiss temperature, and $\chiO$ a temperature-independent paramagnetic susceptibility. The fit parameters in both cases are reported in Table~\ref{tab:MCW}. The evaluated effective magnetic moments per Eu atom $\meff$ have slightly larger values for measurements in magnetic fields applied parallel to the $\cax$ than for measurements with $\Hperpc$. This was observed also in other \Eu-based systems.\cite{Tran2012NJP,Tran2012PRB,Jiang2009NJP,Jiang2009PRB,Zapf2017} The determined $\meff\sim\unit[8]{\muB}$
is marginally larger than
$\meff^{\text{theo}}=g\sqrt{J(J+1)}=\unit[7.9]{\muB}$ expected from theory, which is also consistent with previously reported results\citep{Zapf2017} and is usually  attributed to a possible contribution of $\Feion$ magnetic moments.
The constant $\chiO$ contribution is negligibly small when compared to the main magnetic susceptibility, confirming the absence of spurious phases.

\begin{table}[h]
\begin{centering}
\caption{Magnetic parameters of \EuCa{} and~\EuCaCa{}, as evaluated from fitting the
temperature-dependent susceptibility investigated in $\Hperpc$ and $\Hparc$
using Eq.~\ref{eq:MCW}.\label{tab:MCW}}

\par\end{centering}
\centering{}%
\begin{ruledtabular}
\begin{tabular}{lcccc}
 & \multicolumn{2}{c}{$\EuCa$} & \multicolumn{2}{c}{$\EuCaCa$}\\
Orientation & $\Hperpc$ & $\Hparc$ & $\Hperpc$ & $\Hparc$\\
\hline
$\meff$ $\left(\muB\right)$ & $8.1$ & $8.2$ & $7.9$ & $8.2$\\
$\unit[\Thp]{(K)}$ & $17.1$ & $13.9$ & $11.9$ & $7.8$\\
$\unit[\chiO]{\left(cm^3/Eu\right)}$ & $0.005$ & --- & $0.025$ & $0.008$\\
\end{tabular}
\end{ruledtabular}
\end{table}

Positive Weiss temperatures $\Thp$ suggest dominant ferromagnetic interactions in both compounds. 
Similar results were obtained in 
\Eu{}\cite{Jiang2009NJP} and \ce{EuCo2As2},\cite{Ballinger2012} too, 
most likely 
indicating also there the presence of
ferromagnetic interactions between nearest-neighbor \Euion{} ions in the $\abplane$s.

\subsection{ac susceptibility\label{ssec:ACsusc}}

%
\begin{figure}[ht]
\includegraphics[width=0.49\textwidth]{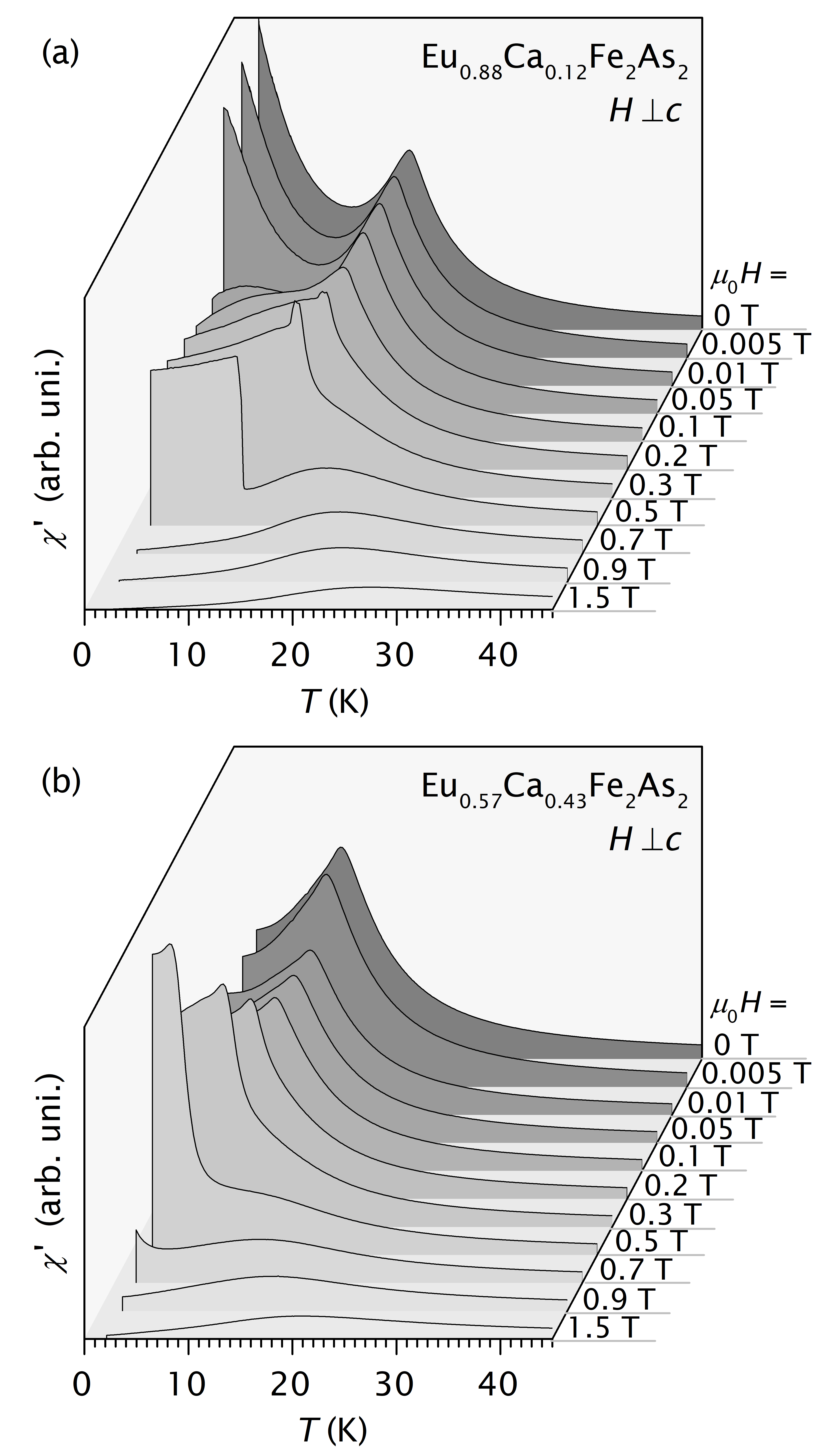}
\caption{Temperature dependencies of the real-part of ac susceptibility $\chi'$
of (a) $\EuCa$ and (b) $\EuCaCa$ measured in several external magnetic
fields ($\muO H$) applied perpendicular to the crystallographic $\cax$.\label{fig:ac}}
\end{figure}
%

The temperature dependence of the real part of ac susceptibility $\chi'$
of \EuCa{} and \EuCaCa{} measured at selected $\Hperpc$ fields is shown
in Fig.~\ref{fig:ac}. The zero-field data, show a rather wide peak associated with the AF ordering of \Euion{} ions (at $\TN\approx\unit[15.6]{K}$ for \EuCa{} and at \unit[10]{K} for \EuCaCa{}). As expected for an AF transition, the peak shifts towards lower temperatures
as the external magnetic field increases.

%
\begin{figure}[ht]
\includegraphics[width=0.49\textwidth]{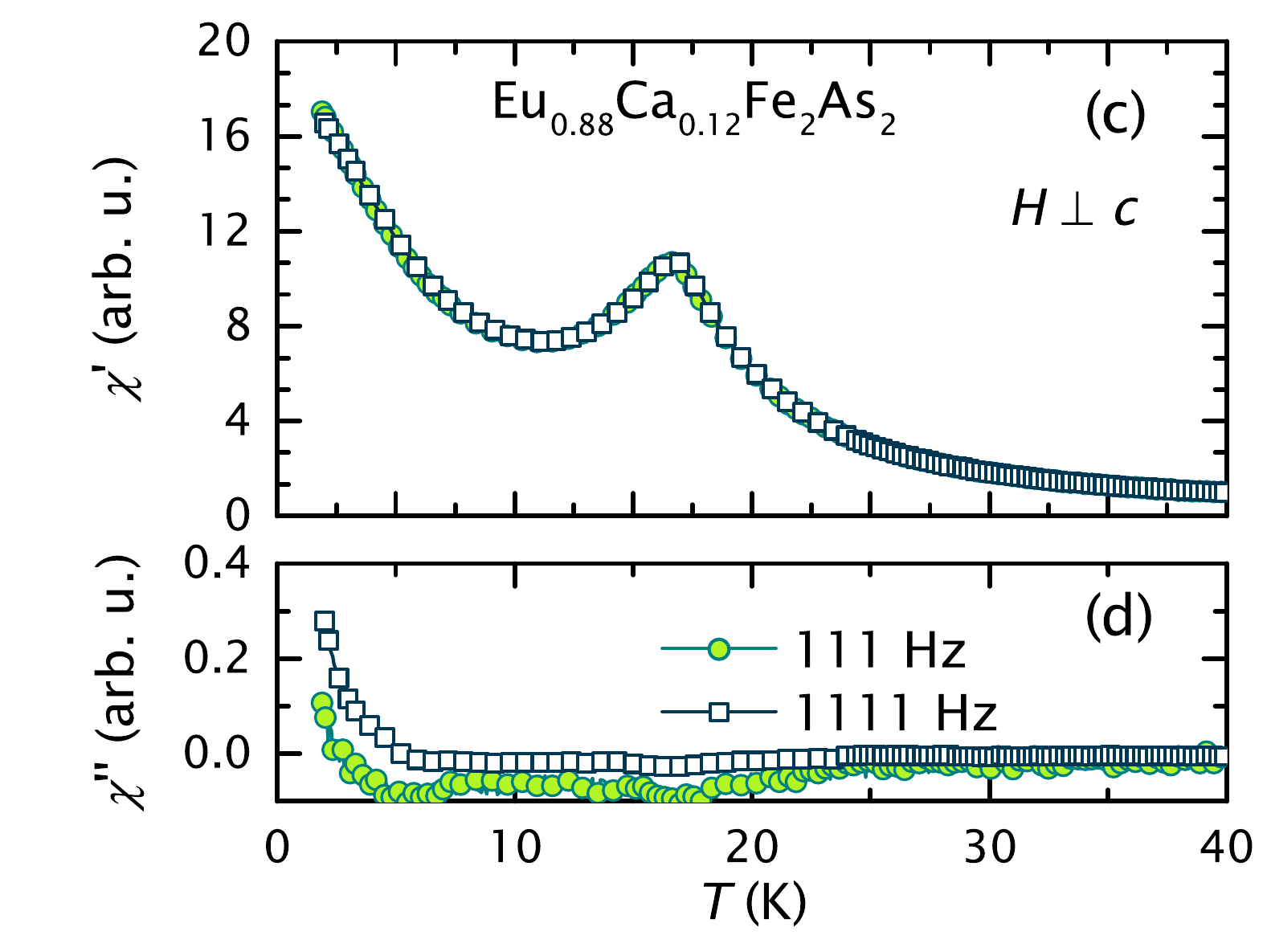}
\caption{Temperature dependence of (a) real and (b) imaginary parts of ac susceptibility of \EuCa{} measured in zero external magnetic field, using a driving field of $\muO\Hac=\unit[10]{mT}$ ($\perp~c$) with frequencies of $\unit[111]{Hz}$ and $\unit[1.111]{kHz}$.\label{fig:ac-im}}
\end{figure}
%

At temperatures below $\TN$ and for fields below $\unit[0.1]{T}$ both the real and imaginary part of the \EuCa{} susceptibility increase (see Fig.~\ref{fig:ac-im}). Such behavior in a magnetically-ordered state suggests the existence
of additional phase transitions, especially since
in the case of the P\nobreakdash-doped \Eu-based compounds a spin-glass transition was observed in an already magnetically ordered state.\cite{Zapf2013}

The zero-field ($\muO H = \unit[0]{T}$) ac-susceptibility data collected
at different frequencies reveal that although the imaginary part increases slightly at higher frequencies, the real part invariably shows
the same temperature dependence (cf. Fig.~\ref{fig:ac-im}). Additionally, constructed $\chi''(\chi')$ plots (see Supplementary Material) for investigated compounds do not show a semicircular shape expected for a spin-glass system,\cite{Mydosh,Petracic2003} hence ruling out a spin-glass transition at low temperatures.

Taking into account the magnetization measurement results (see Sec.~\ref{ssec:Mag}), we propose that the observed increase of ac susceptibility at low temperatures is possibly associated with the  ``detwinning'' processes.
Such suggestion is further supported by the ac-susceptibility measurements of
the \Eu{} parent compound (with an A-type magnetic structure),
which reveal exactly the same behavior
as in the 12\% Ca-doped compound, i.e.,
a significant increase in the real and imaginary parts of ac susceptibility below the AF transition (see Supplementary Material for more details).

On the other hand, such behavior was not observed in Co-doped compounds with a canted-AF structure.\cite{Tran2012NJP,Tran2012PRB}

Hence, we propose that the observed increase of susceptibility with decreasing temperature below $\TN$ in both \Eu{} and \EuCa{} is associated with the ``detwinning'' processes, indicated by the magnetization measurements.

\subsection{Magnetic phase diagrams\label{ssec:PhDiag}}
\begin{figure*}[t]
\centering
\includegraphics[width=0.95\textwidth]{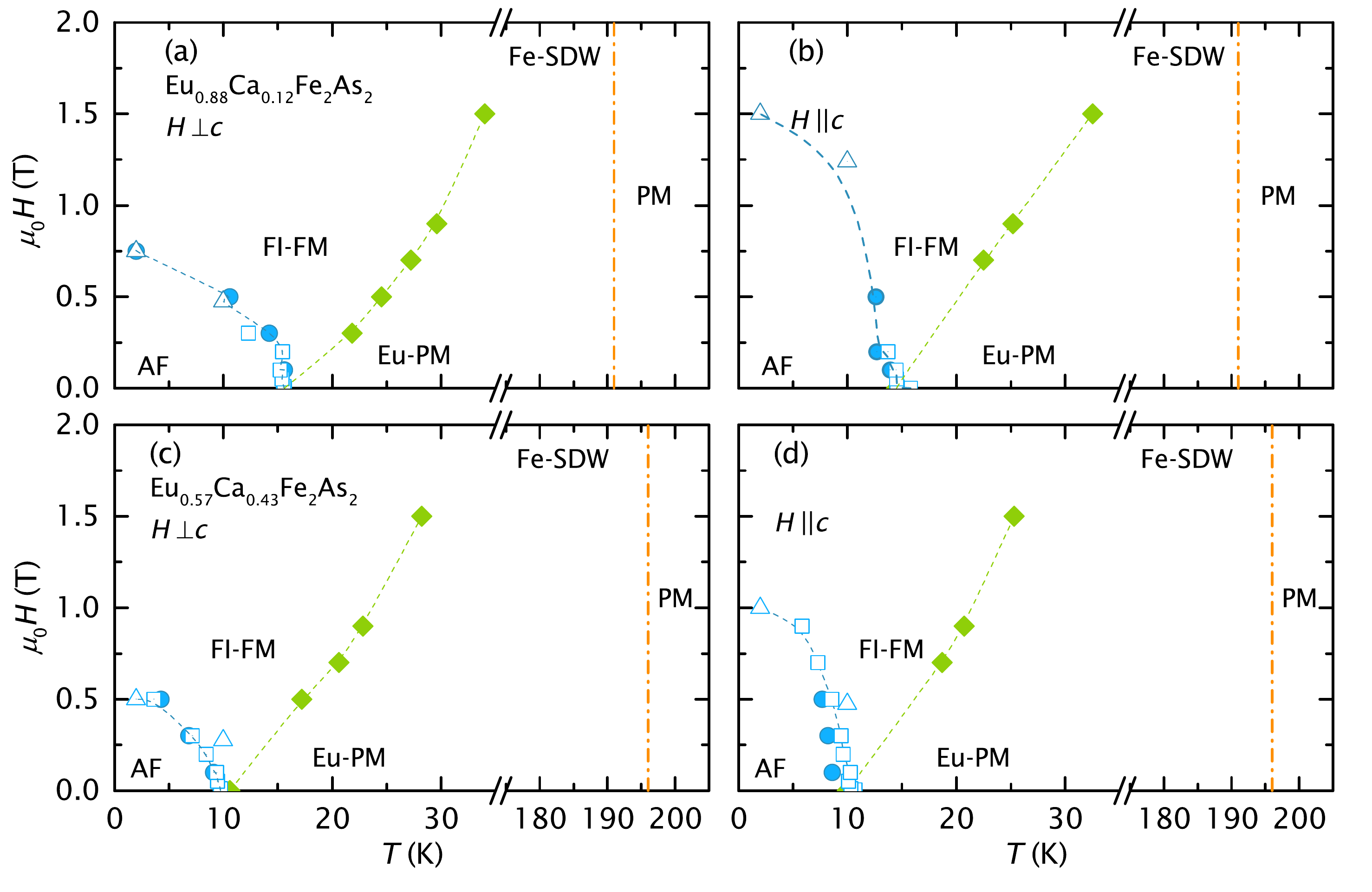}
\caption{Magnetic phase diagrams of (a,b - top panels) \EuCa{} and (c,d - bottom panels) \EuCaCa{} in
magnetic fields applied (a,d) parallel and (b,d) perpendicular to the $\cax$. Dashed lines are guides for the eye.\label{fig:diag}}
\end{figure*}
%

Based on the above bulk measurements of \EuCa{} and \EuCaCa{},
the respective magnetic-phase diagrams were constructed for
fields applied both parallel and perpendicular to the $\cax$ (see Fig.~\ref{fig:diag}). The vertical dashed-dotted lines represent the transition from the paramagnetic (PM) to the Fe-SDW transition. The curves corresponding to the AF to FI-FM transition, related to the \Euion{} magnetic moments, were constructed by taking the $\Hcr$ (triangles) and $\TN$ (circles) values, calculated from the dc magnetization data, and the $\TN$ (squares) from ac susceptibility.
The transition from the paramagnetic phase of the Eu-subsystem (Eu-PM) to the ordered FI-FM phase (diamonds)
was determined based on the ac-susceptibility values.
The dashed lines on the diagrams represent guides for the eye.

The presented phase diagrams reveal a rather anisotropic behavior. 
Thus, at \unit[2]{K}, in both cases the $\Hcr$ values for $H\parallel c$  (Fig.~\ref{fig:diag} - right column) are almost twice
the $\Hcr$ values obtained for $H\perp c$ (Fig.~\ref{fig:diag} - left column). By comparing instead the $\Hcr$ values of the two compounds (top vs.\ bottom pannels in Fig.~\ref{fig:diag})
we deduce that the RKKY interactions become weaker upon doping, since smaller magnetic fields have to be applied
to reorient the nearly 50\%-doped system from the AF to the FI-FM state with respect to the 12\% Ca-doping case.

\subsection{Muon-spin relaxation\label{ssec:musr}}

Spin-polarized muons are widely used microscopic as probes of magnetism.\cite{Blundell1999,Yaouanc2011}
Once implanted in matter and thermalized at interstitial lattice sites, they act as
probes of the local magnetic field, the muon-spin precession frequency being
$\nu = \gamma_{\mu}/(2\pi) B_{\mathrm{loc}}$, with
$\gamma_{\mu} = 2\pi \times 135.53$\,MHz/T, the muon gyromagnetic ratio.
Since the final signal is carried by energetic decay positrons (emitted preferentially
along the muon-spin direction), $\mu$SR enables the investigation of materials
even in the absence of applied fields.

\begin{figure}[ht]
  \centering
  \includegraphics[width=0.45\textwidth]{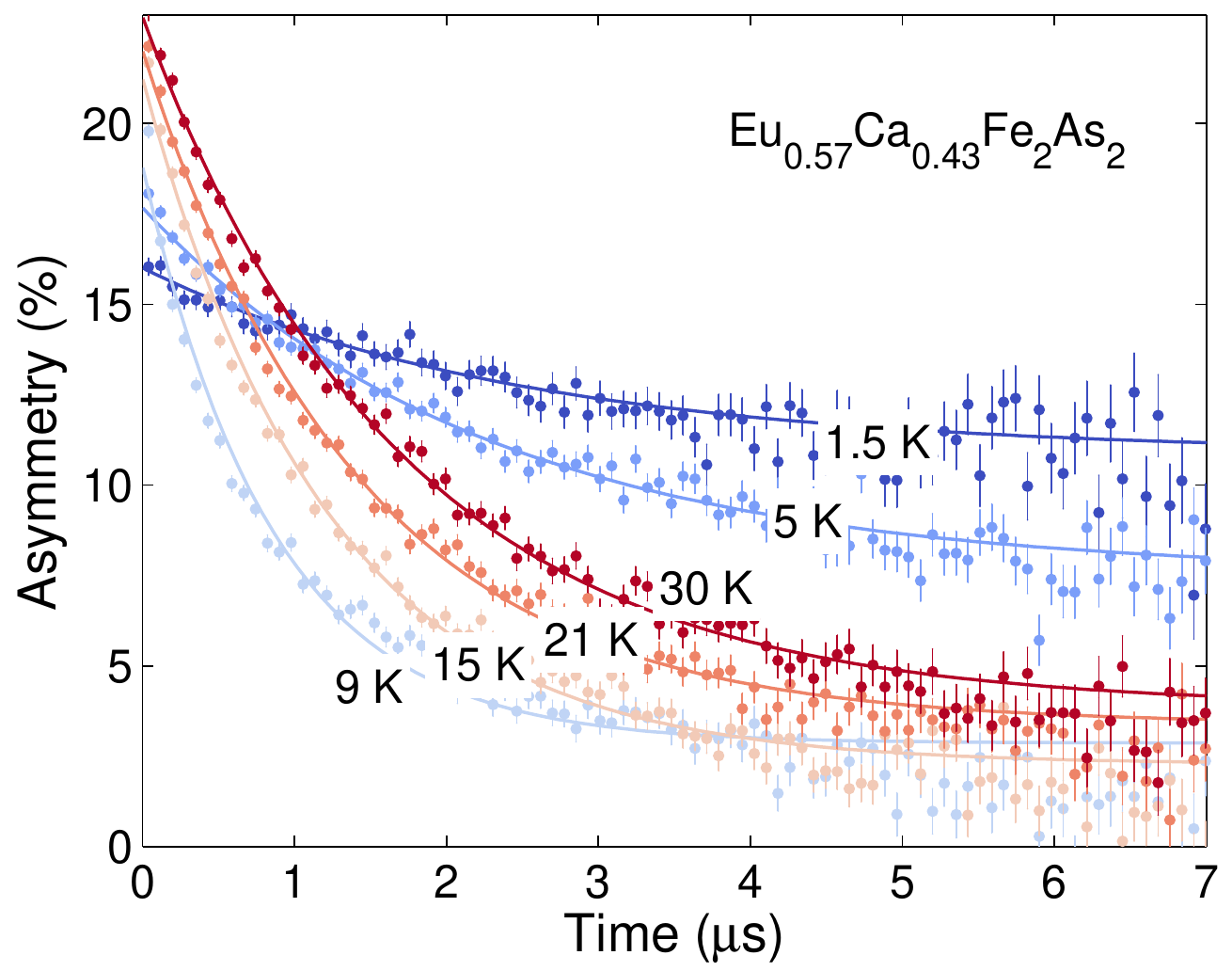}   
  \caption{\label{fig:MuSR_spectra}Time-domain zero-field 
  $\mu$SR asymmetry spectra of Eu$_{1-x}$Ca$_x$Fe$_2$As$_2$ for
  $x= 0.43$ at selected temperatures observed on a long time scale.
  Solid curves represent fits by
  means of the relaxation function (\ref{eq:MuSR_relax}).}
\end{figure}
%

%
\begin{figure}[ht]
\centering
\includegraphics[width=0.238\textwidth,angle=0]{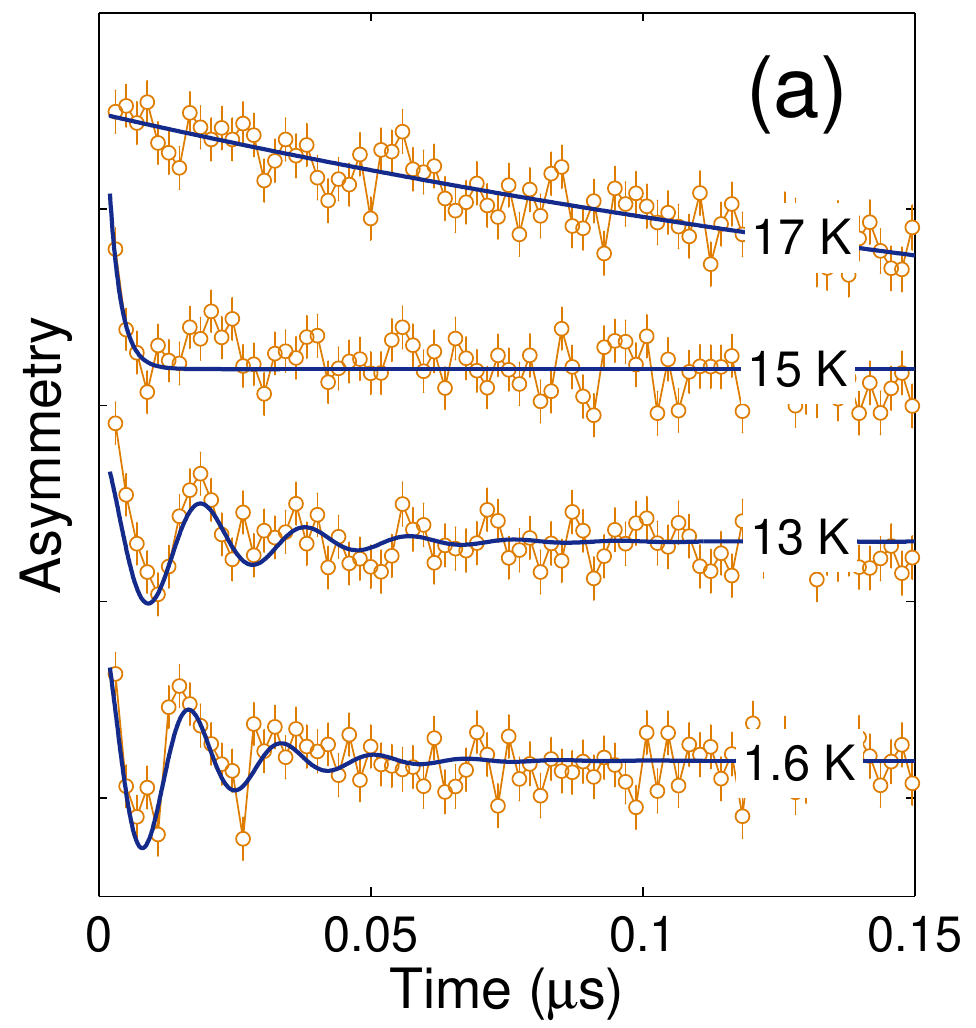}
\includegraphics[width=0.237\textwidth,angle=0]{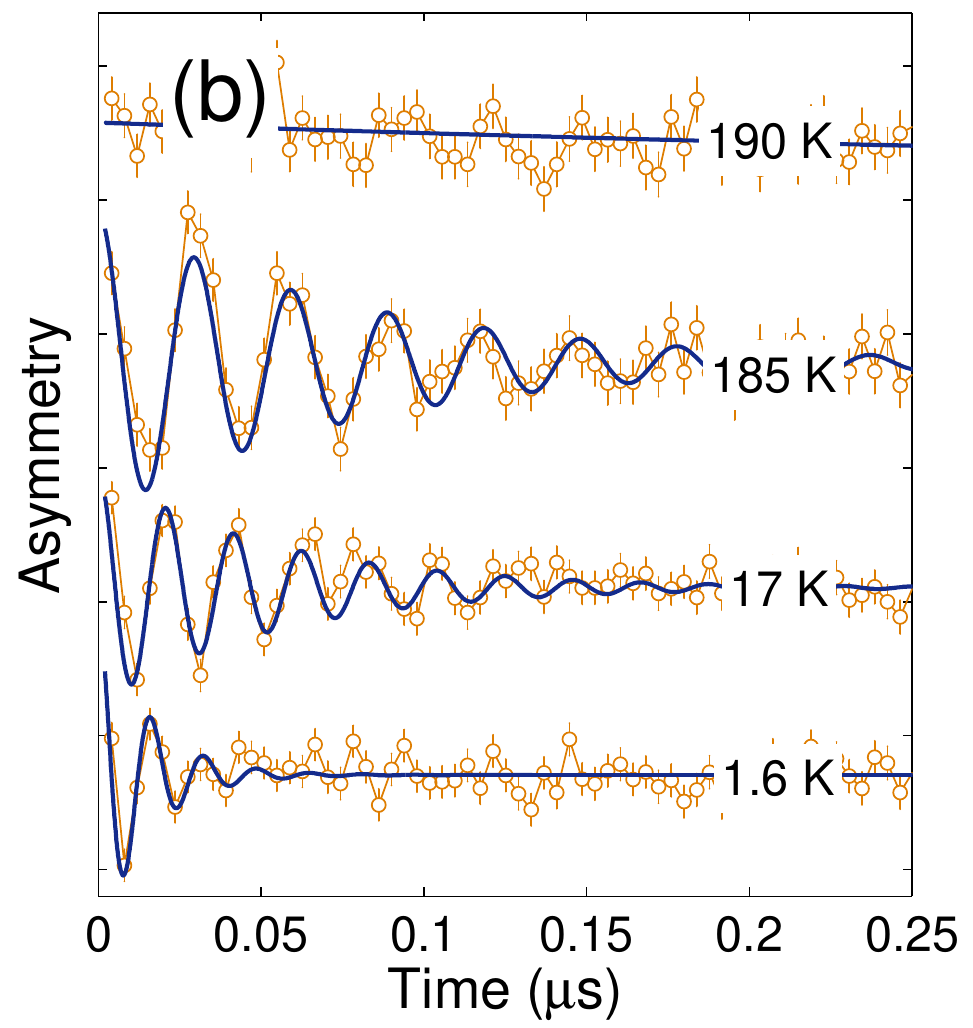}
\vspace{-2ex}%
\caption{\label{fig:short_time_comp}Comparison of the short-time ZF-$\mu$SR
asymmetries for $x = 0.12$, as measured in the FB (left) and UD (right) detectors.
While in the first case the oscillations disappear above the $T_\mathrm{N}^\mathrm{Eu}$,
in the second case, they persist up to the $T_\mathrm{N}^\mathrm{Fe}$.
The observed spectra are compatible with the iron and europium magnetic
moments lying in the $ab$-plane (see text).
}
\end{figure}
%

To investigate the magnetic behavior of the Ca-doped EuFe$_2$As$_2$
system with temperature, 
muons with spins parallel to the $\cax$ were implanted into the
samples with $x(\mathrm{Ca}) = 0.12$ and 0.43, representative of a
weak and strong dilution of the Eu$^{2+}$ magnetic lattice, respectively.
Typical time-domain asymmetry spectra for the $x = 0.43$ case, collected upon warming
the sample, are shown in Fig.~\ref{fig:MuSR_spectra}.
At first sight, no asymmetry oscillations are observed on a long time scale
at any of the temperatures shown, either above or below $T_{\mathrm{N}} = 11$\,K.
Except for a higher $T_{\mathrm{N}}$, similar results are
obtained also in the less diluted $x = 0.12$ case.
In both cases, such an \emph{apparent} lack of oscillations
on either side of $T_{\mathrm{N}}$ is due to rather different reasons.
For temperatures above $T_{\mathrm{N}}$, a smooth decay starting with a
full initial asymmetry (of 22\% at 30\,K) corresponds to the typical
behavior of fluctuating paramagnetic moments (Eu$^{2+}$ in our case).
Below  $T_{\mathrm{N}}$ instead, a significant drop in the initial asymmetry
(reaching 16\% at 1.5\,K) reflects the onset of a long-range magnetic order.
By limiting the observation time window to 0.2\,$\mu$s, one can clearly see
the oscillations in the magnetically ordered phase [see Fig.~\ref{fig:short_time_comp}(a)].
The rather large Eu$^{2+}$ magnetic moments ($\sim 7$\,$\mu_\mathrm{B}$)
imply significant internal fields (above 0.4\,T) and, hence,
fast precessions, which then disappear above $T_\mathrm{N}^\mathrm{Eu}$.

\begin{figure}[ht!]
\includegraphics[width=0.3\textwidth]{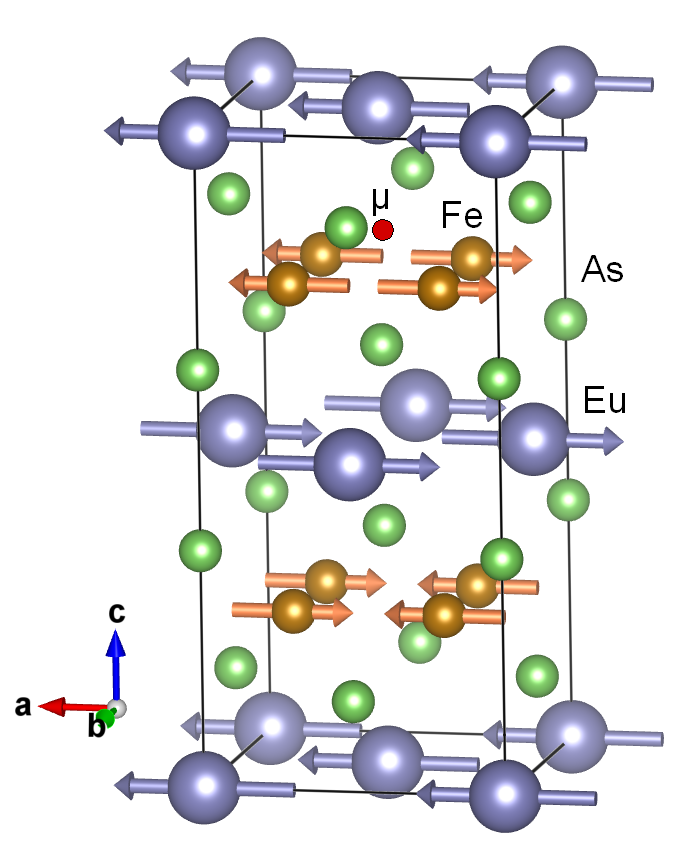} 
\caption{\label{fig:magn_struct}The magnetic structure of Ca-doped
EuFe$_2$As$_2$, as inferred from $\mu$SR data, is most likely the same as
that of the \Eu{} parent compound. The red circle shows a possible
(high-symmetry) muon stopping site. Drawing produced using VESTA.\cite{Vesta3}
}
\end{figure}
%

%
\begin{figure}[ht]
\centering
\includegraphics[width=0.4\textwidth,angle=0]{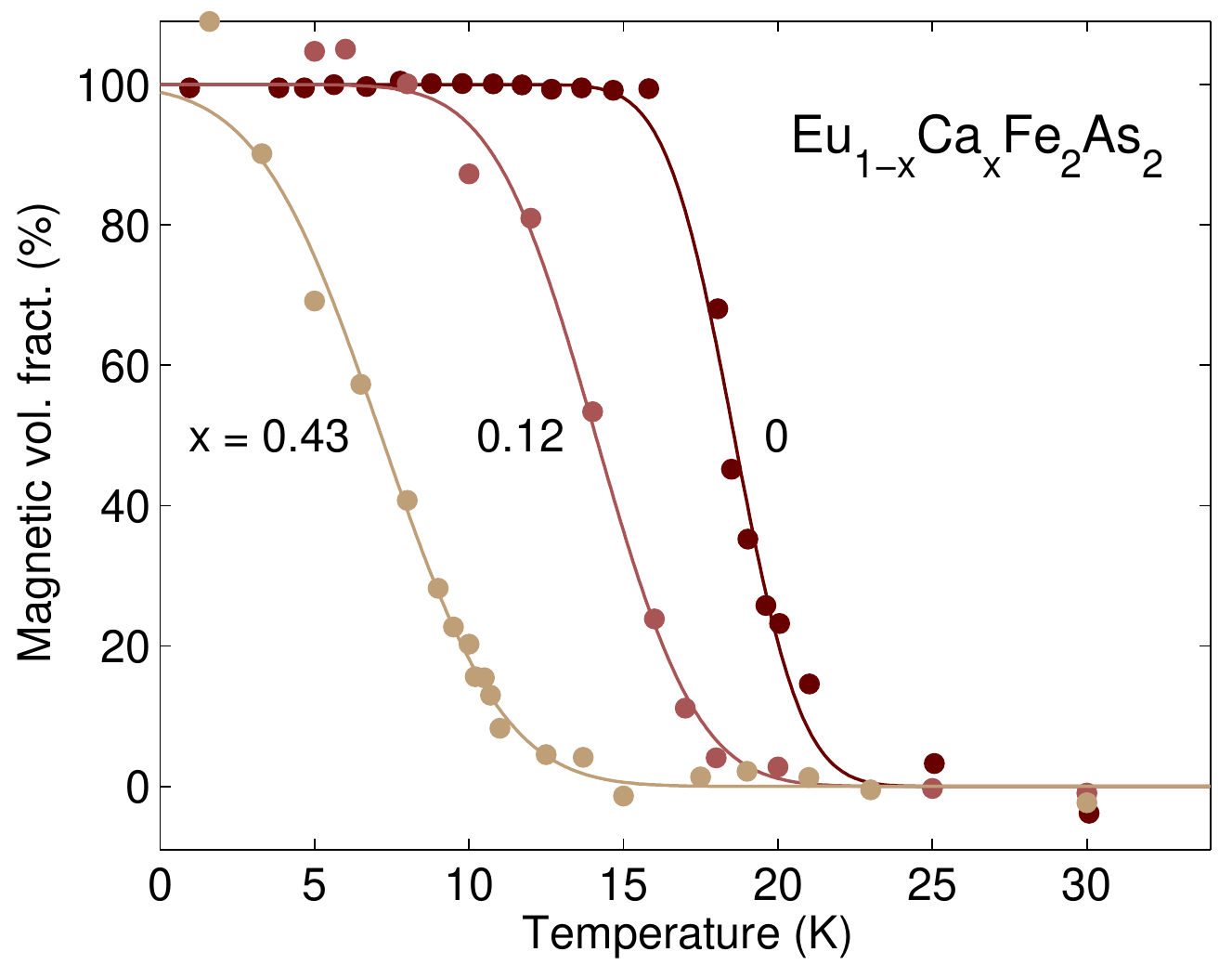}  
\vspace{-2ex}%
\caption{\label{fig:Magn_vol_fract_Eu} Magnetic volume fraction vs.\
temperature as extracted from the longitudinal component of $\mu$SR data
and fitted with an $\mathrm{erf}(T)$ function [Eq.~(\ref{eq:erf})] for 
samples with $x(\mathrm{Ca}) = 0$, 0.12, and 0.43. The data for the
$x = 0$ case were obtained from Ref.~\onlinecite{Anand2015}.}
\end{figure}
%

A comparison of the forward-backward (FB) vs.\ up-down (UD) detector pairs
(Fig.~\ref{fig:short_time_comp}) is quite informative. Given the single-crystal
nature of our samples, it allows us to infer also the orientation of the
Eu$^{2+}$ and Fe$^{2+}$ magnetic moments in their respective ordered phases.
Actually the $\mu$SR signal can only give indications about the local
magnetic field at the muon site, since it does not contain direct information
on the magnetic moment directions. However, by taking into account the
high symmetry of the muon stopping site [most likely close to the Fe planes
at ($1/2$, $1/2$, 0.22) --- see, e.g., Ref.~\onlinecite{Maeter2009} and
Fig.~\ref{fig:magn_struct}], one can still draw conclusions on the direction
of magnetic moments for both Fe$^{2+}$ and Eu$^{2+}$.

Since incoming muons have spins parallel to the $\cax$, a lack of
oscillations above $T_\mathrm{N}^\mathrm{Eu}$ in the FB pair, while
they still persist in the UD detectors, implies that in the long-range ordered
SDW phase the iron moments align in the $ab$-plane.
Indeed, each muon has four nearest Fe$^{2+}$ ions which, considering the
$\boldsymbol{k} =  (1, 0 , 1)$ SDW propagation vector,\cite{Xiao2009} implies two
groups of counteraligned moments $\boldsymbol{m}$ (see Fig.~\ref{fig:magn_struct}).
For symmetry reasons, the contributions to the local field at the muon site due to the
$c$ component of  $\boldsymbol{m}$ cancel out, whereas the in-plane component gives
rise to a local dipolar field along the $c$-axis. By observing (below $T_\mathrm{SDW}$)
a local field along $c$, we therefore infer that the Fe$^{2+}$ moments should
lie in the $ab$-plane.

As for the europium moments, since $T_\mathrm{N}^\mathrm{Eu}$ is lower than
$T_\mathrm{N}^\mathrm{Fe}$, one cannot unambigously determine their orientations,
since below $T_\mathrm{N}^\mathrm{Eu}$ muons are simultaneously affected by both
magnetic systems. However, compatibly with the data shown in
Fig.~\ref{fig:short_time_comp}(a), we expect Eu$^{2+}$ moments most
likely to lie in the $\abplane$. Indeed, each muon has a single nearest Eu$^{2+}$ neighbor,
shifted along the $c$-direction. If the Eu$^{2+}$ moment would lie along the $c$-direction,
the dipolar field at the muon site would be antiparallel to the moment and along $c$
(which is not our case). If the Eu$^{2+}$ moments would lie in the $ab$-plane, the
dipolar field would still be antiparallel, but now in the $ab$-plane. Since below $T_\mathrm{N}$
we observe a local field in the $ab$-plane, we conclude that the Eu$^{2+}$ magnetic
moments lie also in the $ab$-plane.
The above conclusions are compatible with neutron diffraction results on
single crystals of EuFe$_{2}$As$_{2}$,
where both Eu$^{2+}$ and Fe$^{2+}$ moments are shown to align along the
$a$-axis (at 2.5\,K).

From the above discussion (and a comparison of Figs.~\ref{fig:MuSR_spectra}
and \ref{fig:short_time_comp}), it is clear that the considered time-scale
determines the choice of a suitable fitting function: oscillatory for
short times and slowly decaying at long times.
The typical time-domain $\mu$SR spectra reported in Fig.~\ref{fig:MuSR_spectra}
show a strongly temperature-dependent decay which was fitted
using the following combined 
function:
\begin{equation}
\label{eq:MuSR_relax}
A(t) = A_0 \left[\alpha e^{-\lambda_{T}t}\cos(\gamma_{\mu}B_{i}t + \phi) + \beta
e^{-\lambda_{L}t} \right].
\end{equation}
Here $\alpha$ and $\beta =1-\alpha$ are the oscillating (i.e.,
transverse) and non-oscillating (i.e., longitudinal) fractions of the
muon signal, respectively, whose initial total asymmetry is $A_0$.
Accordingly, the two decay functions are denoted as $\lambda_{T}$ and
$\lambda_{L}$. The inital precession angle is denoted by $\phi$, while
$\gamma_{\mu}$ is the muon gyromagnetic ratio.
For the ideal case of a fully magnetic polycrystalline sample one expects
$\alpha = 2/3$ and $\beta = 1/3$ (since statistically one third of the
times the muon spin does not precess, being parallel to the local
magnetic field $B_{i}$). Similar values for the two fractions have been
found also for single crystals,\cite{Guguchia2013} or for partially
aligned mosaics (as in our case --- see discussion below).

As already mentioned, 
the apparent reduction of the longitudinal asymmetry $\beta A_{0}$
(obtained by considering the long-time $\mu$SR spectra --- see Fig.~\ref{fig:MuSR_spectra})
is a consequence of the onset of the Eu$^{2+}$ magnetic
order below $T_\mathrm{N}$. The full recovery of asymmetry to its
total value of ca.\ 25\% occurs
only in the paramagnetic phase (i.e., above $T_{c} \sim 190$\,K).
However, data show that most of the asymmetry is recovered already above
$T_\mathrm{N}$. Considering this and the rather similar oscillating
fractions $\alpha$ in a powder and in a mosaic sample, one can attempt
an evaluation of the magnetic fraction $V_{M}$ as a function of doping
and temperature by means of
$V_{M}(T) = \nicefrac{3}{2}\,(1-a_{L}) \times 100\%$.\cite{Shiroka2011}
The resulting $V_{M}(T)$ values for the pristine and Ca-substituted
EuFe$_{2}$As$_{2}$ samples are shown in Fig.~\ref{fig:Magn_vol_fract_Eu}.

%
\begin{figure}[ht]
\centering
\includegraphics[width=0.38\textwidth,angle=0]{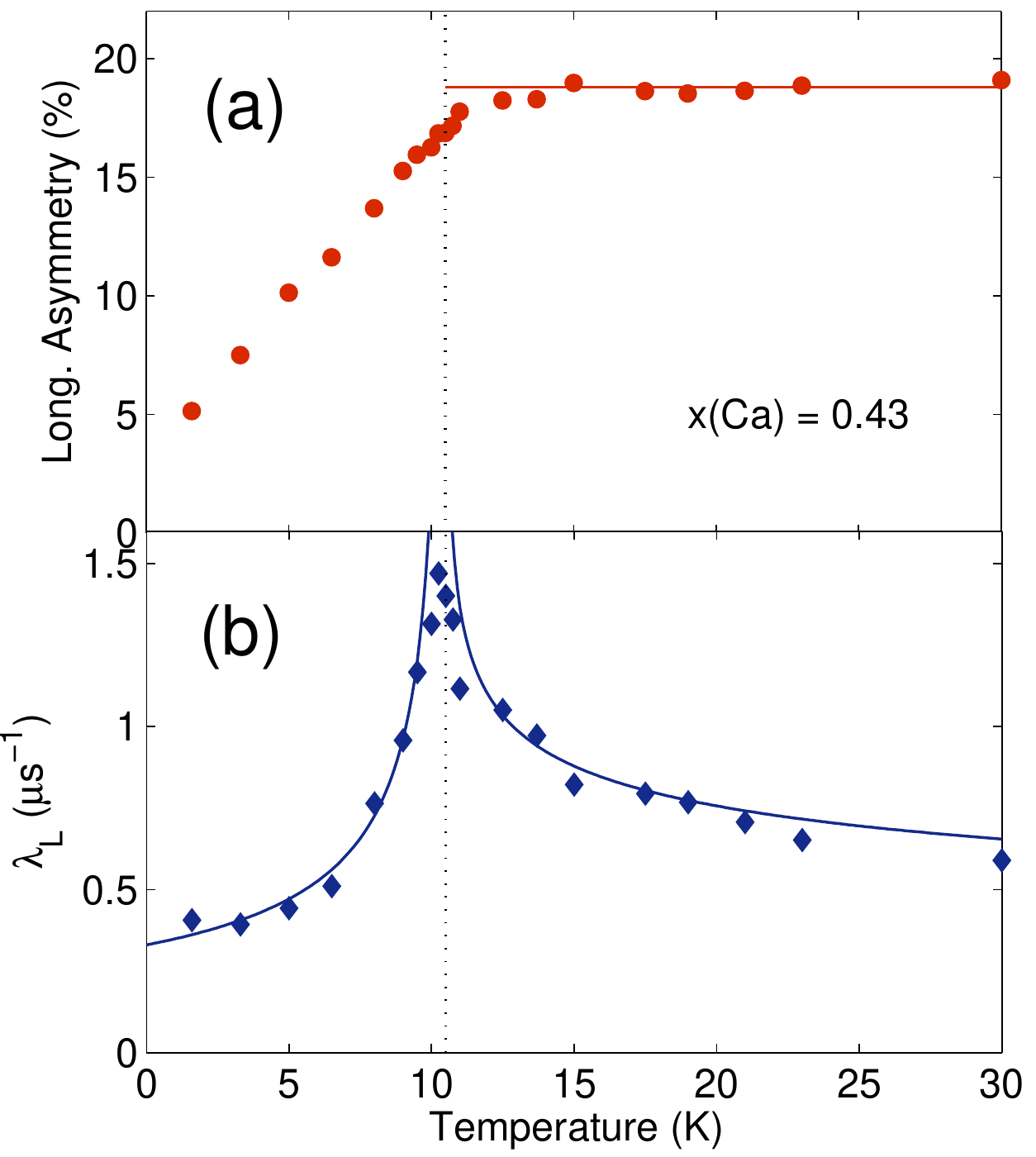}  
\vspace{-2ex}%
\caption{\label{fig:ZF_MuSR_Ca1}Fit parameters of the long-time $\mu$SR
spectra vs.\ temperature for the $x = 0.43$ case: (a) 
longitudinal muon-spin asymmetry at 0 and 200\,mT, (b) 
longitudinal decay coefficient. Both parameters show marked features (slope change or cusp)
in concomitance with the Eu$^{2+}$ AF phase transition at $T_\mathrm{N} = 11$\,K.}
\end{figure}
%

We note that all the considered samples are fully magnetically ordered
at low temperature ($V_{M} = 100\%$). The average N\'eel temperature and
the corresponding transition width $\Delta$ were obtained by fitting
the $V_{M}(T)$ data with the phenomenological function:\cite{Shiroka2011}
\begin{equation}
\label{eq:erf}
V_{M}(T) = \frac{1}{2}\,\left[1 - \mathrm{erf}\left(\frac{T - T_{\mathrm{N}}}%
{\sqrt{2}\Delta}\right)\right].
\end{equation}
%
As can be seen from Fig.~\ref{fig:Magn_vol_fract_Eu}, as the Ca content
increases there is a clear decrease of $T_\mathrm{N}$ and a simultaneous
broadening of the transition (the curves becomes smoother). These results are
in very good agreement with our magnetometry data. In addition, they show
that to a high Ca substitution rate corresponds an increased degree of
disorder, here measured by $\Delta$. The detailed fit values are reported
in Table~\ref{tab:fit_erf}.

\begin{table}[h]
\centering
\renewcommand{\arraystretch}{1.2}
\caption{\label{tab:fit_erf}Magnetic properties of
Eu$_{1-x}$Ca$_{x}$Fe$_{2}$As$_{2}$ samples as a function of Ca substitution.}
\begin{ruledtabular}
\begin{tabular}{p{1mm}lcccc}
\vspace{2mm}
& \lower 0.7mm \hbox{$x$(Ca)}
& \lower 0.7mm \hbox{$T_{\mathrm{N}}$ (K)}
& \lower 0.7mm \hbox{$\Delta$ (K)}
& \lower 0.7mm \hbox{Magn. frac.\ (\%)}\\[2pt]
\hline
&0        &18.6(1) &1.7(3)  &  99(4) \\
&0.12(1)  &15.1(3) &2.6(3)  & 100(5) \\
&0.43(1)  &7.2(4)  &3.1(4)  & 100(8) \\
\end{tabular}
\end{ruledtabular}
\end{table}
%

\begin{figure}[ht]
\centering
\includegraphics[width=0.4\textwidth,angle=0]{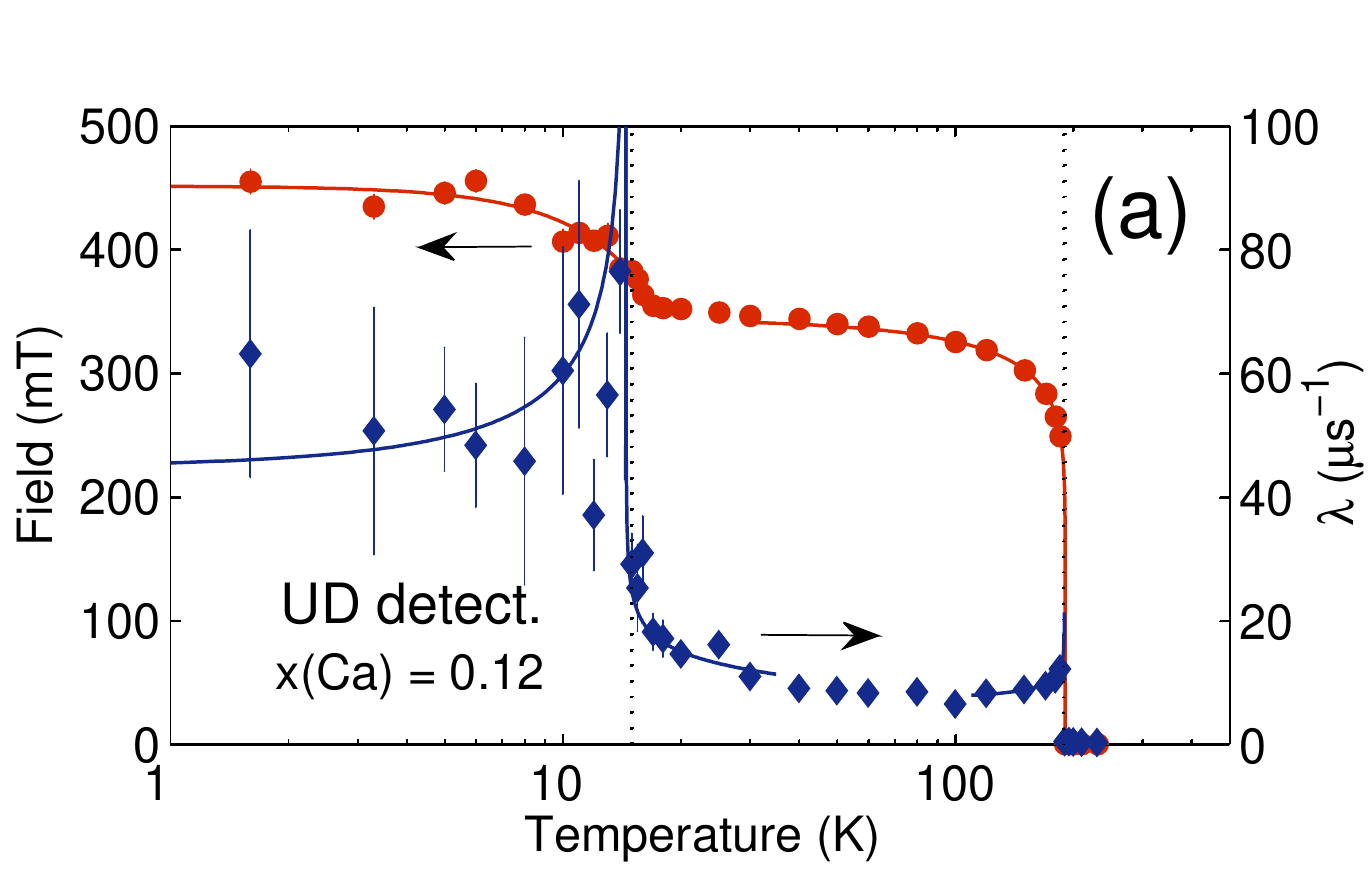} 
\includegraphics[width=0.41\textwidth,angle=0]{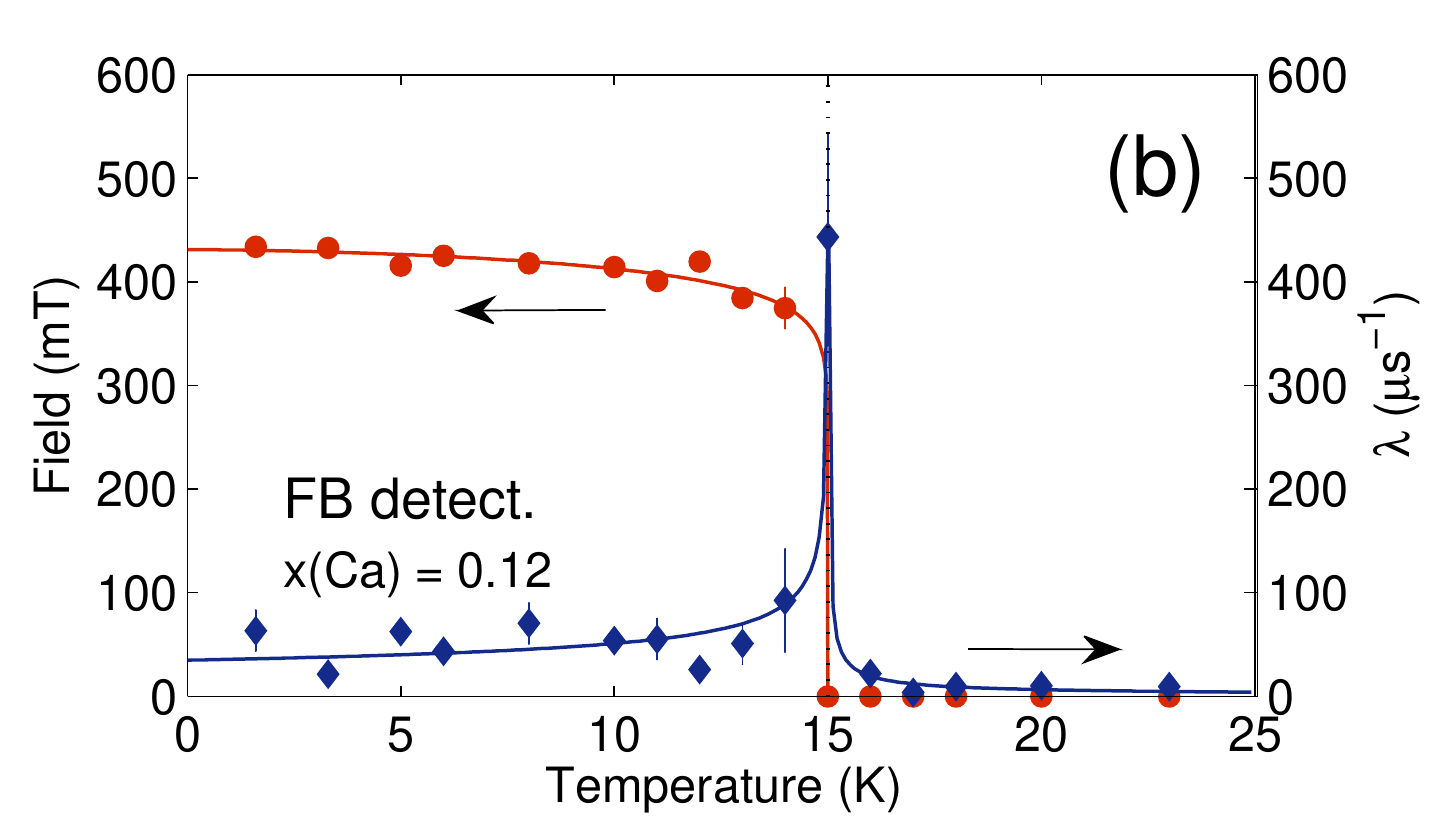} 
\vspace{-2ex}%
\caption{\label{fig:overview}(a) Short-time ZF-$\mu$SR fit parameters 
vs.\ temperature for the $x = 0.12$ case, showing the Eu- and Fe-related phase
transitions at 15 and 190\,K, respectively (dashed lines). Since data refer to the
UD detector pair, both transitions are observed. The internal magnetic field
(circles), was determined from the $\mu$SR oscillation frequency.
The decay rate (diamonds) shows marked peaks in concomitance with
the two magnetic phase transitions.
(b) Same as in (a), but data refer to the FB detector pair.
The observed field (left scale) goes abruptly to zero at $T_\mathrm{N}^\mathrm{Eu}$,
where the decay rate (right scale) exhibits a very marked divergence.
}
\end{figure}

By considering the fit parameter $\lambda_{L}$, although strictly speaking
the relaxation rate 
is not an order parameter, it still
provides information on the critical dynamics, as suggested by the sharp cusps 
observed at $T_\mathrm{N}$
(diamond symbols in Figs.~\ref{fig:ZF_MuSR_Ca1} and \ref{fig:overview}).
The $T$ dependence of $\lambda_{L}$, both above and below $T_\mathrm{N}$,
has been described by means of a critical-exponent behavior
$\lambda(T) = \lambda_{0}\left|1 - T/T_\mathrm{N}\right|^{-w}$,\cite{Yaouanc1993,Henneberger1999}
with the expected exponent value for a 3D ferromagnet $w \approx 1.05$.
The fit curves, shown with solid lines in Fig.~\ref{fig:ZF_MuSR_Ca1}(b),
have  rather different exponents above and below $T_\mathrm{N}$.
Thus for $T < T_\mathrm{N}$ we find $w \simeq 0.84(3)$, whereas for
$T > T_\mathrm{N}$ we obtain $w \simeq 0.25(2)$. While the former value is
not far from that expected at a magnetic transition, the latter is much
smaller and remains so even when restricting the fit range. Such
anomalously small $w$ values 
above $T_\mathrm{N}$ have been observed also elsewhere (see, e.g.,
Ref.~\onlinecite{Anand2015}), where $w$ was found to be 0.28.
It has been suggested that muon-lattice dipolar interactions, known to strongly
affect the paramagnetic critical dynamics near a magnetic transition
point, could be responsible for the observed low value of the $w$
exponent.\cite{Yaouanc1993,Henneberger1999}


%
%
%
%
%

%
By considering now the short-time $\mu$SR data (see Fig.~\ref{fig:short_time_comp})
we can follow the temperature evolution of the internal magnetic field,
as sensed at the muon stopping sites. The resulting fit parameters in
the $x = 0.12$ case, obtained by means of Eq.~(\ref{eq:MuSR_relax}),
are shown in Figs.~\ref{fig:overview}(a) and (b) for the UD and FB
detector pairs, respectively. In comparison with the longitudinal
relaxation data [see Fig.~\ref{fig:ZF_MuSR_Ca1}(b)] the transverse relaxation
is more than two orders(!) of magnitude larger.
Similarly large relaxation rates have been observed also in the
isoelectronically doped EuFe$_{2}$(As$_{1-x}$P$_{x}$)$_{2}$
compound.\cite{Guguchia2013}

As discussed above, the UD detector dataset allow us to follow both
the Eu- and Fe-related magnetic transitions, distinctly seen at
$T_\mathrm{N} = 15$\,K and $T_\mathrm{SDW} = 190$\,K, respectively, in
the field profiles and in the muon relaxation rates [see
Fig.~\ref{fig:overview}(a)]. By converse, the FB detector dataset is limited
to the Eu$^{2+}$ transition only.
Analogously to the longitudinal relaxation, also the transverse relaxation
data show a critical-exponent behavior %
$\lambda(T) = \lambda_{0}\left|1 - T/T_\mathrm{N}\right|^{-w}$,
once more with rather skewed exponent values above $T_\mathrm{N}$.

As for the internal field dependence with $T$, the very sharp increase of
$B_{i}$ at $T_\mathrm{SDW}$ is suggestive of a first-order phase transition.
As the temperature is lowered, the internal field tends to saturate at
ca.\ 0.35\,T. At $T_\mathrm{N}$ another field jump appears, indicative of 
the magnetic ordering of Eu$^{2+}$ moments. Each of these transitions can be
modelled by the phenomenological model:\cite{Le1993,Blundell1995}
\begin{equation}
\label{eq:field_fit}
B_{i}(T) = \sum_{n=1}^{2}\,B_{i}^{n}(0)\,\left[1 - \left(\frac{T}{T_{c}^{n}}\right)^{\gamma_n}\right]^{\delta_n},
\end{equation}
where $B_{i}^{n}(0)$ are the internal magnetic fields at zero temperature
and $\gamma_n$ and $\delta_n$ are two empirical parameters, whose
typical fit values are $\gamma_n \sim 1.5$ and $\delta_n \sim 0.25$
(with error bars of ca.\ 0.8).

For both samples, none of the internal magnetic fields seems to track an
ideal mean-field type curve, which represents the exact solution in
case of an infinite-range (anti)ferromagnetic order and corresponds to
$\delta_n \sim 0.5$.
The reason for such discrepancy might reflect the occurrence of a magnetic
(AF) order within an already ordered (SDW) phase.

\section{Summary and conclusions\label{sec:Conclusions}}

Our study suggests that similarly to the \Eu{} parent compound, \EuCa{} (and most likely \EuCaCa{})
undergo analogous 
``detwinning'' processes. Some differences observed between the two Ca-doped compounds
are possibly due to the different coupling strengths between the Fe- and Eu-subsystems, associated
with the different Eu-concentration in the two cases.

The $\mu$SR results offer a local-probe perspective on the Ca-doped \Eu{} system, e.g.,
implanted muons indicate that in the magnetically-ordered phase both the \Feion{}
and \Euion{} magnetic moments align in the $ab$-plane.
The latter is also supported by magnetometry results, which suggest an A-type AF structure
for the \Euion{} magnetic moments in the Ca-doped compounds, the same as in the \Eu{} case.
At low temperatures, both samples show 100\% magnetic fractions, but a dilution
of the Eu$^{2+}$ sublattice implies a decrease of $T_\mathrm{N}$ and a broadening
of the transition. In both cases, muons detect magnetic fields of ca.\ $450$\,mT
in the ordered phase, with the muon field profiles and relaxation rates showing
distinct features at the respective phase transitions of the two magnetic sublattices.
The presence of an AF order, even in the nearly 50\% diluted system,
excludes direct interactions among Eu$^{2+}$ moments as responsible
for the observed 3D magnetic order and suggests instead the indirect RKKY
couplings to be the key interactions.
Such conclusion is supported also by theoretical calculations,
as shown, e.g., in Ref.~\onlinecite{Akbari2011}.
The significant reduction of $\TN$ and $\Hcr$ in EuFe$_{2}$As$_{2}$
upon Eu-by-Ca substitution indicates a clear weakening of the
RKKY-mediated magnetic interactions among Eu$^{2+}$ moments. 
This result could be rationalized by considering the direct influence of local on-site
substitutions and is in clear contrast with the outcome of substitutions
in the FeAs-planes (such as, e.g., Co-for-Fe), for which the temperature of the
AF ordering of Eu$^{2+}$ moments remains mostly unchanged upon doping.\cite{Guguchia2011}

Finally, we observe also a small change of $\TSDW$ values upon doping, which might suggest
some degree of interaction between the Eu and Fe layers. This is consistent
with theoretical predictions, which indicate that the Eu order might be influenced by the Fe-magnetism.\cite{Akbari2011,Maiwald2018} However,
one cannot exclude that 
a $\TSDW$ change might 
arise simply because of 
a change in the lattice parameters or other processes. Therefore, the investigation
of similar compounds but with smaller or (preferably) with no changes of lattice parameters,
will be helpful to solve this issue.

Given the interesting zero-pressure results reported above, future high-pressure
$\mu$SR investigations of Ca-doped Eu\-Fe$_2$\-Ca$_2$ are promising,
since the Ca-doped EuFe$_2$As$_2$ system is expected to become a superconductor above 1.5--2\,GPa.

\begin{acknowledgments}
We thank H.\ Luetkens and J.\ Barker
for the experimental support at PSI.
This work is partially based on experiments performed at the Swiss Muon Source
S$\mu$S, Paul Scherrer Institute, Villigen, Switzerland and was financially
supported in part by the Schweizerische Nationalfonds zur F\"{o}rderung der
Wissenschaftlichen Forschung (SNF).
\end{acknowledgments}


%

\end{document}



\title{Supplementary material: Magnetic phase diagram of Ca-substituted \Eu{}}

\author{L.\ M.\ Tran}
\email[Corresponding author: \vspace{8pt}]{l.m.tran@int.pan.wroc.pl}
\affiliation{Institute of Low Temperature and Structure Research, Polish Academy of Sciences, P.O. Box 1410, PL-50-422 Wroclaw, Poland}

\author{M.\ Babij}
\affiliation{Institute of Low Temperature and Structure Research, Polish Academy of Sciences, P.O. Box 1410, PL-50-422 Wroclaw, Poland}

\author{L.\ Korosec}
\affiliation{Laboratorium f\"ur Festk\"orperphysik, ETH Z\"urich, CH-8093 Zurich, Switzerland}

\author{T.\ Shang}
\affiliation{Paul Scherrer Institut, CH-5232 Villigen PSI, Switzerland}
\affiliation{Institute of Condensed Matter Physics, \'Ecole Polytechnique F\'ed\'erale de Lausanne (EPFL), Lausanne CH-1015, Switzerland}

\author{Z.\ Bukowski}
\affiliation{Institute of Low Temperature and Structure Research, Polish Academy of Sciences, P.O. Box 1410, PL-50-422 Wroclaw, Poland}

\author{T.\ Shiroka}
\affiliation{Laboratorium f\"ur Festk\"orperphysik, ETH Z\"urich, CH-8093 Zurich, Switzerland}
\affiliation{Paul Scherrer Institut, CH-5232 Villigen PSI, Switzerland}

\date{\today}

\maketitle

\section{ac susceptibility}

%
\begin{figure*}[h]
\includegraphics[width=0.32\linewidth]{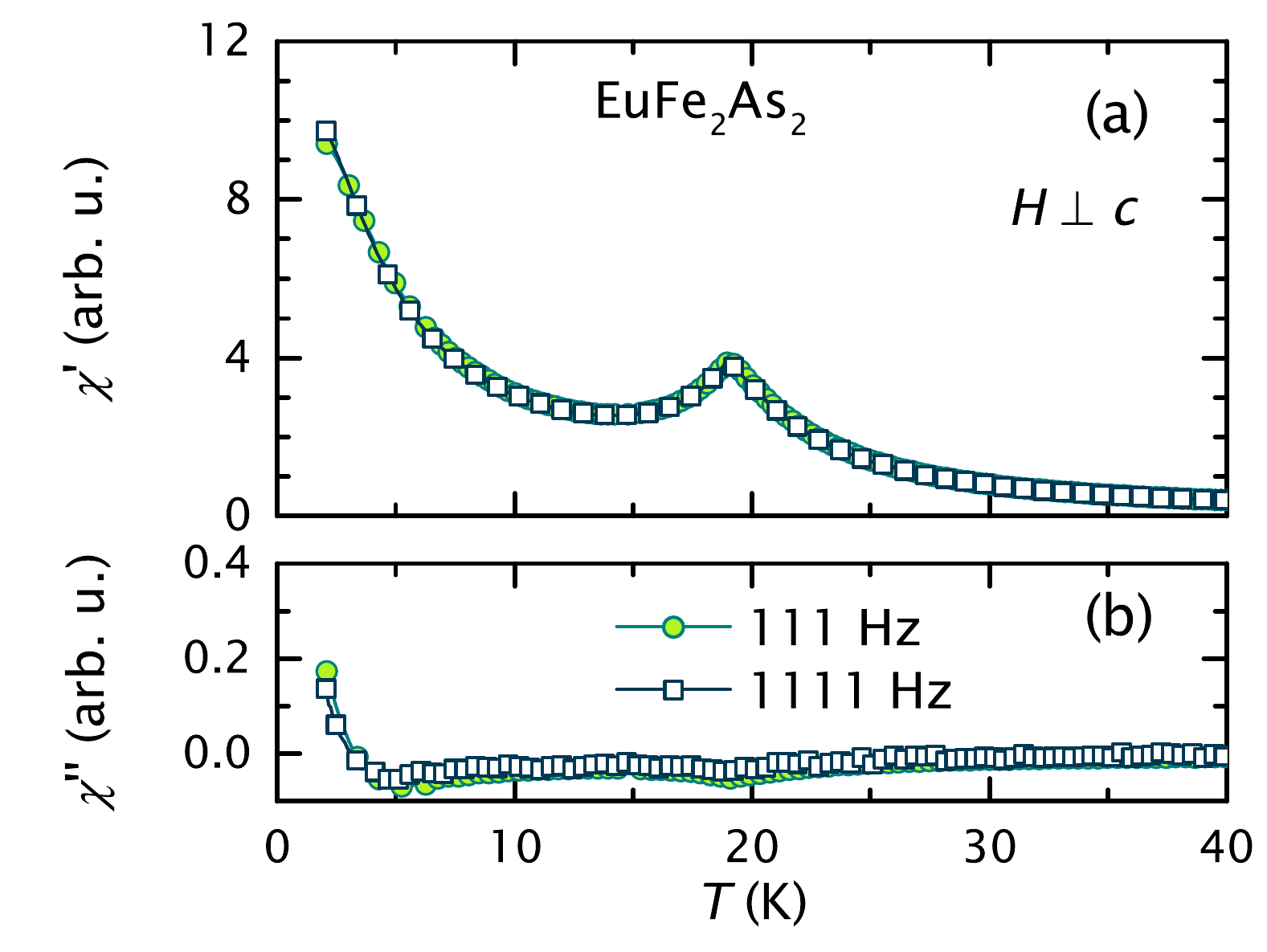}
\includegraphics[width=0.32\linewidth]{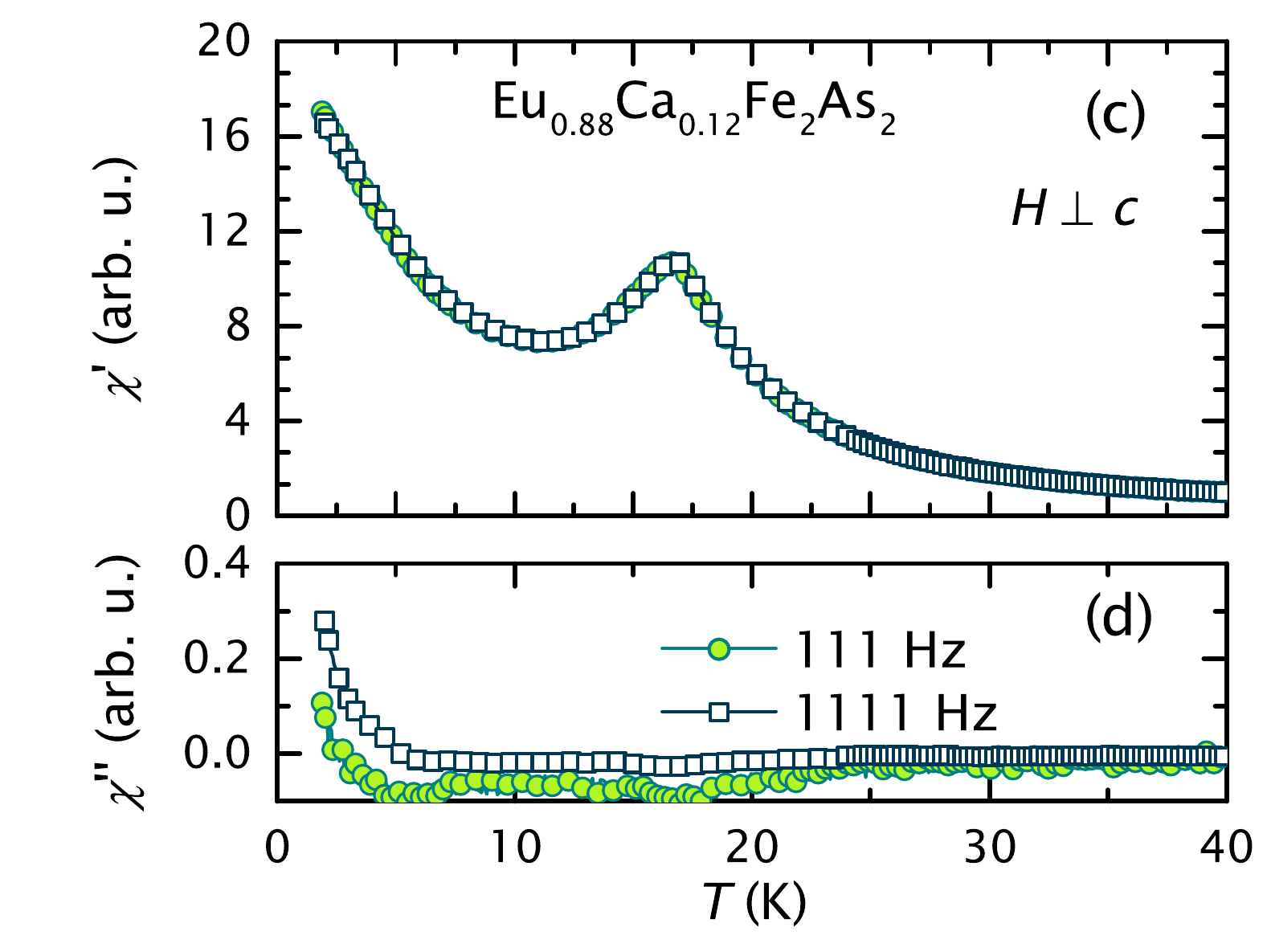}
\includegraphics[width=0.32\linewidth]{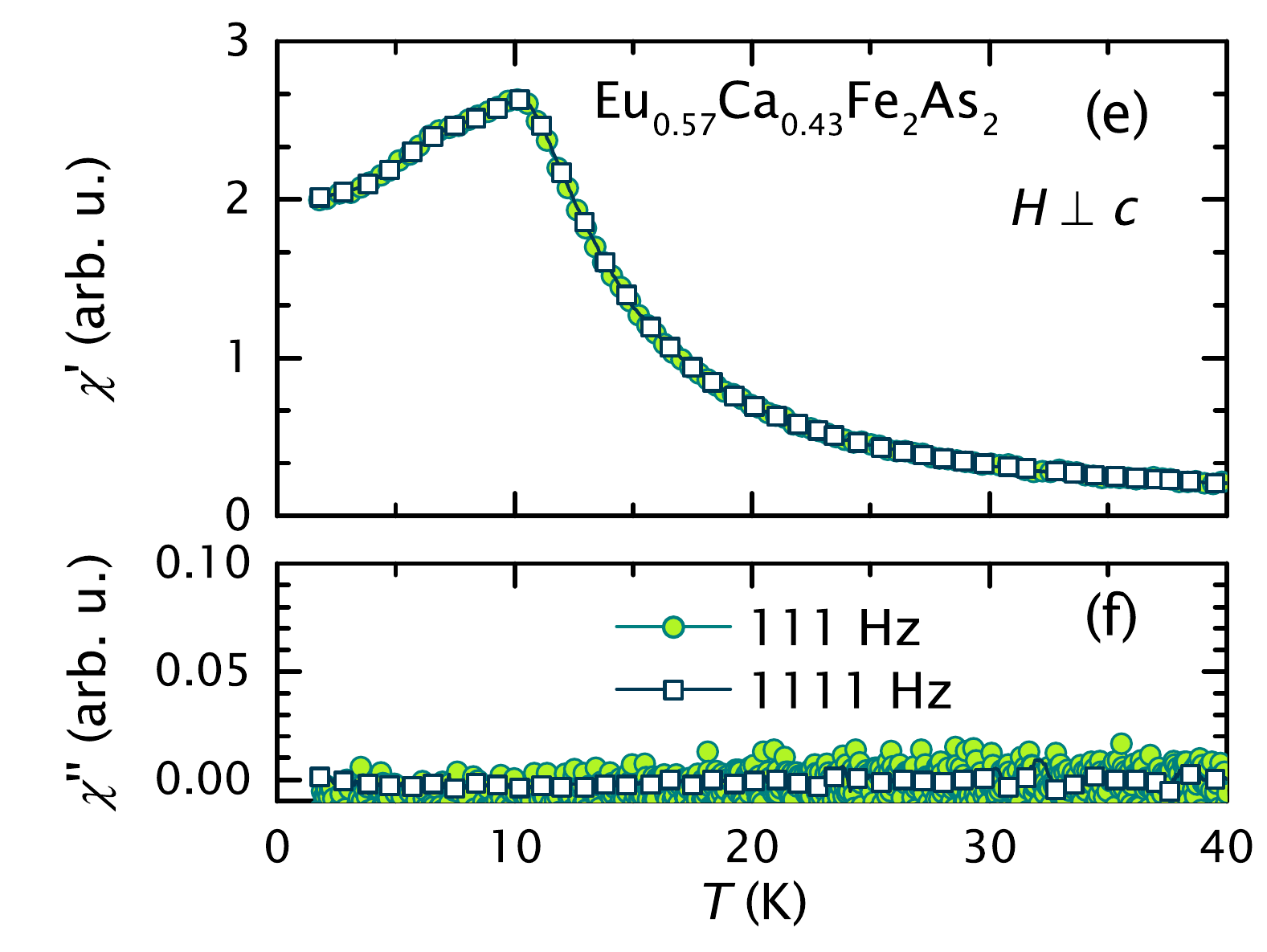}
\caption{Temperature dependence of (a, c, e) real and (b, d, f) imaginar part of ac susceptibility of \EuCaCa{} investigated using 111 and \SI{1111}{\hertz} frequencies of the driving field $\muO\Hac = \SI{10}{\milli\tesla}$\label{fig:ac-im-tot}}
\end{figure*}
%

We investigated the ac susceptibility of \Eu, \EuCa{} and \EuCaCa{} using a driving field of $\muO H = \SI{10}{\milli\tesla}$ with frequencies up to \SI{5}{\kilo\hertz}, in external magnetic fields up to \SI{1.5}{\tesla} applied both parallel and perpendicular to the $\cax$.

The temperature dependencies of zero-field ac susceptibilities of \Eu, \EuCa{} and \EuCaCa{} are presented in Fig.~\ref{fig:ac-im-tot}. All investigated compounds exhibit an anomaly in the real part of ac susceptibility, associated with the antiferromagnetic ordering at $\TN=\SI{19}{\kelvin}$, $\SI{16.5}{\kelvin}$ and \SI{10}{\kelvin} for \Eu{}, \EuCa{} and \EuCaCa{} respectively. For \Eu{} and \EuCa{}, below $\TN$, both the real $\chi'$ and imaginary $\chi''$
parts of ac susceptibility increase with decreasing temperature, while no such behavior was observed for \EuCaCa, cf. Fig.~\ref{fig:ac-im-tot}.

\subsection{Frequency dependence}

To check whether this anomaly could be associated with a spin-glass transition (suggested for the P-doped compounds\cite{Zapf2013}), ac susceptibility was measured for different driving-frequency values.

Measurements performed using frequencies of 111 and \SI{1111}{\hertz} reveal that the temperature dependence of the real part of ac susceptibility for each compound are the same for different frequencies, Fig.~\ref{fig:ac-im-tot}(a,c,e). On the other hand, the imaginary part of ac susceptibility is slightly higher at higher frequencies, Fig.~\ref{fig:ac-im-tot}(b,d,f).

%
\begin{figure*}[h]
\includegraphics[width=0.32\linewidth]{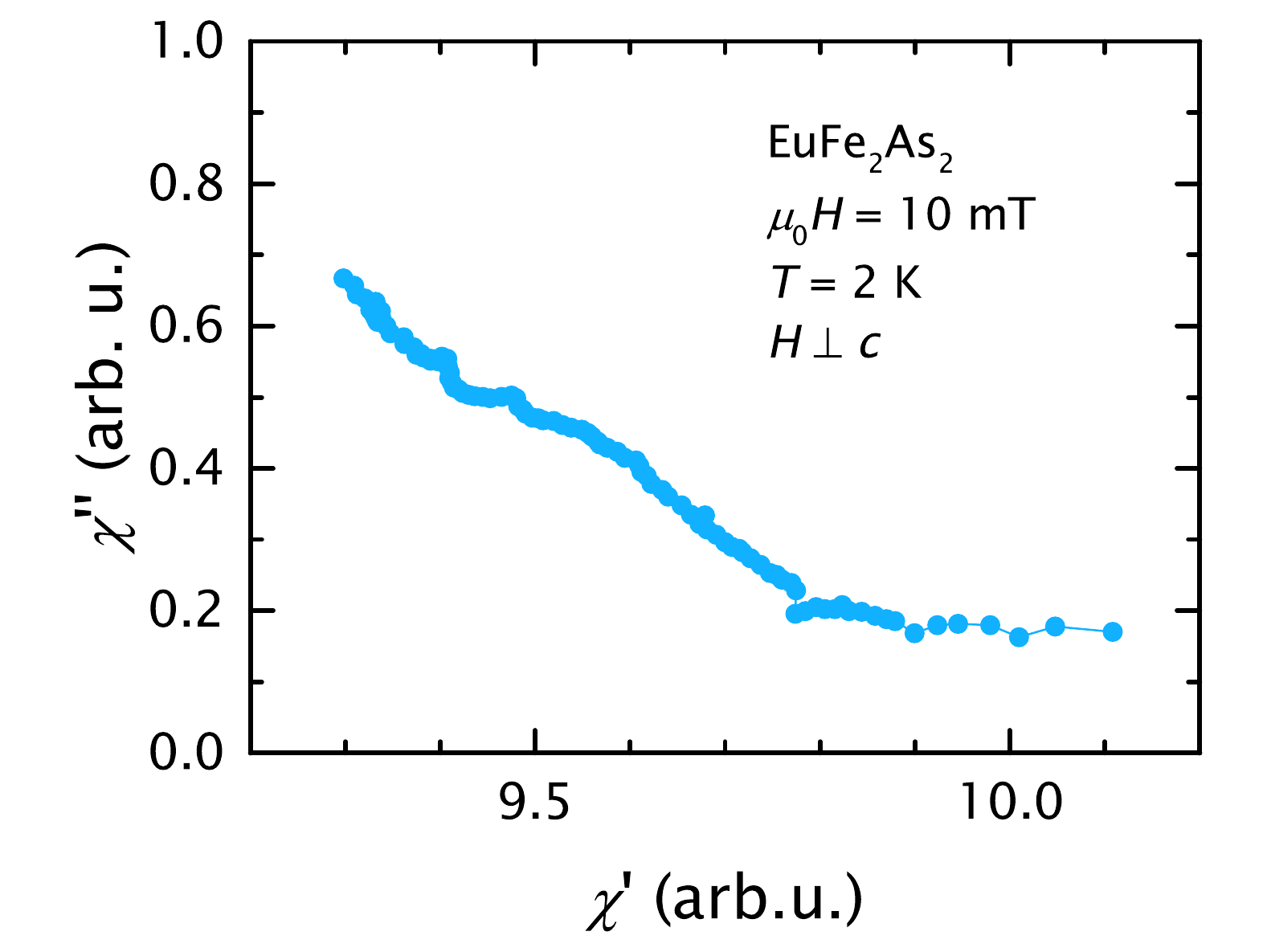}
\includegraphics[width=0.32\linewidth]{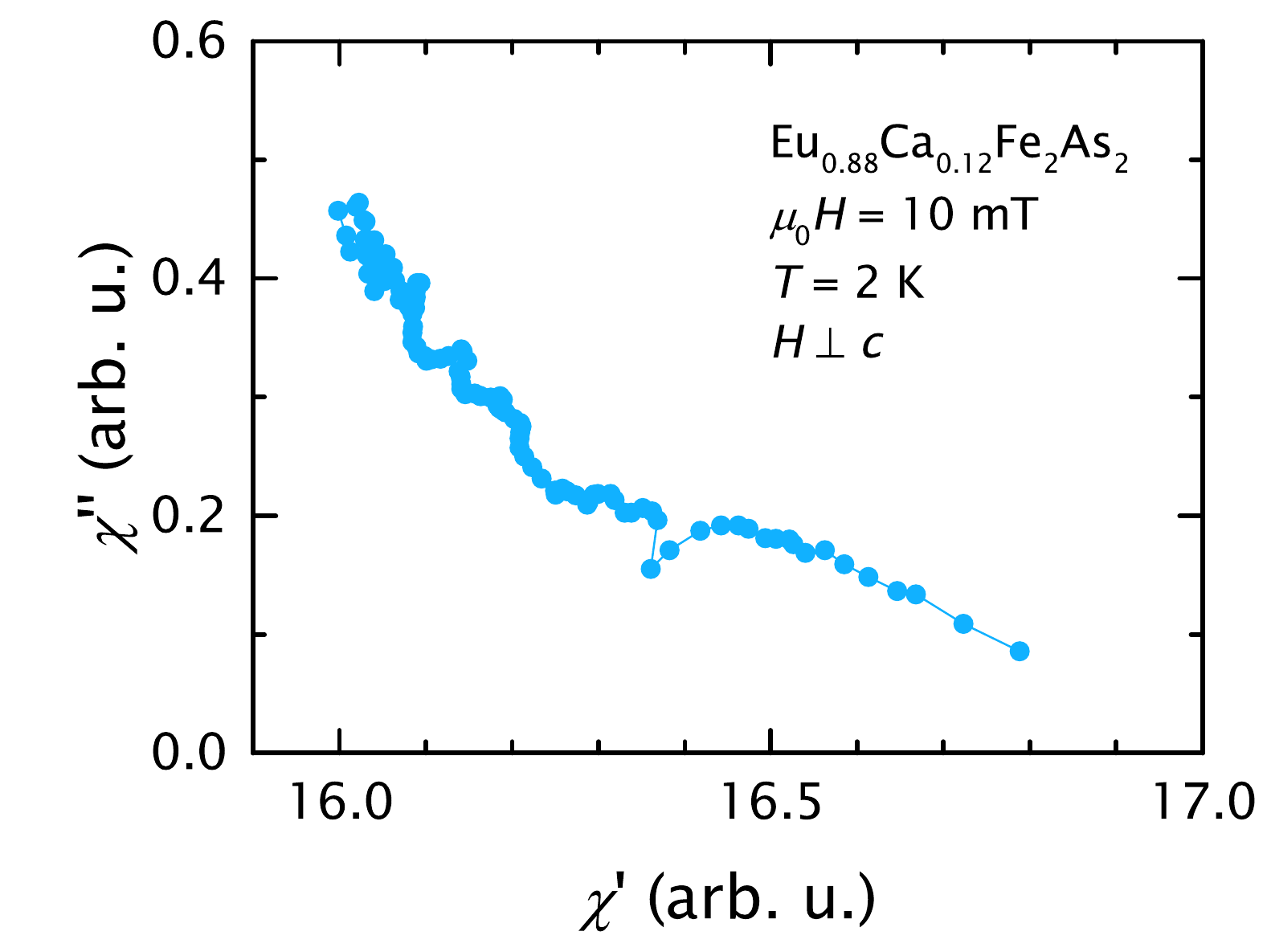}
\includegraphics[width=0.32\textwidth]{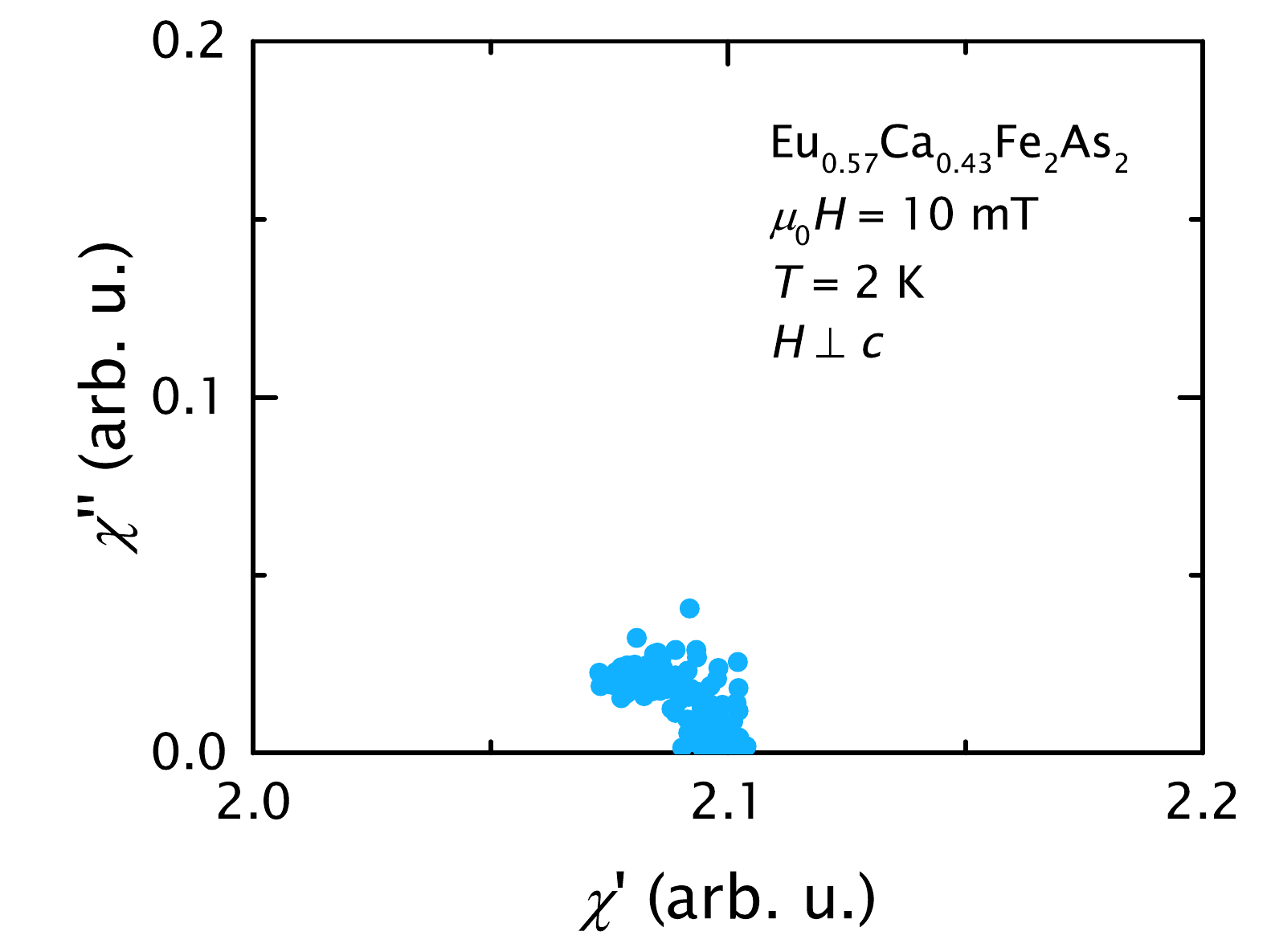}
\caption{$\chi''(\chi')$ plot of \Eu{}, \EuCa{} and \EuCaCa{} investigated at $T=\SI{2}{\kelvin}$ using a driving field $\mu\Hac=\SI{10}{\milli\tesla}$ in the reqency range between 111 and \SI{5500}{Hz}\label{fig:Cole}}
\end{figure*}
%

The frequency dependence of ac susceptibility of \Eu, \EuCa{} and \EuCaCa{} was studied at \SI{2}{\kelvin}. Based on these measurements Argand diagrams\cite{Mydosh} --- $\chi''(\chi')$ plots or Cole-Cole plots --- were constructed 
for each compound and are presented on Fig.~\ref{fig:Cole}. While $\chi''$ increases with increasing frequency, $\chi'$ decreases slightly. The typical $\chi''(\chi')$ dependence for a spin-glass system has a semicircular shape,\cite{Mydosh,Petracic2003} which is not the case for the investigated compounds as can be seen on Fig.~\ref{fig:Cole}. Instead, the $\chi''(\chi')$ dependence is nearly linear. Interestingly, both the temperature dependence and frequency dependence for the nearly 50\% doped compound are qualitatively different compared to the other two compounds.


\subsection{Field dependence}
\begin{figure*}[h]
\hspace{-10pt}\includegraphics[width=0.32\linewidth]{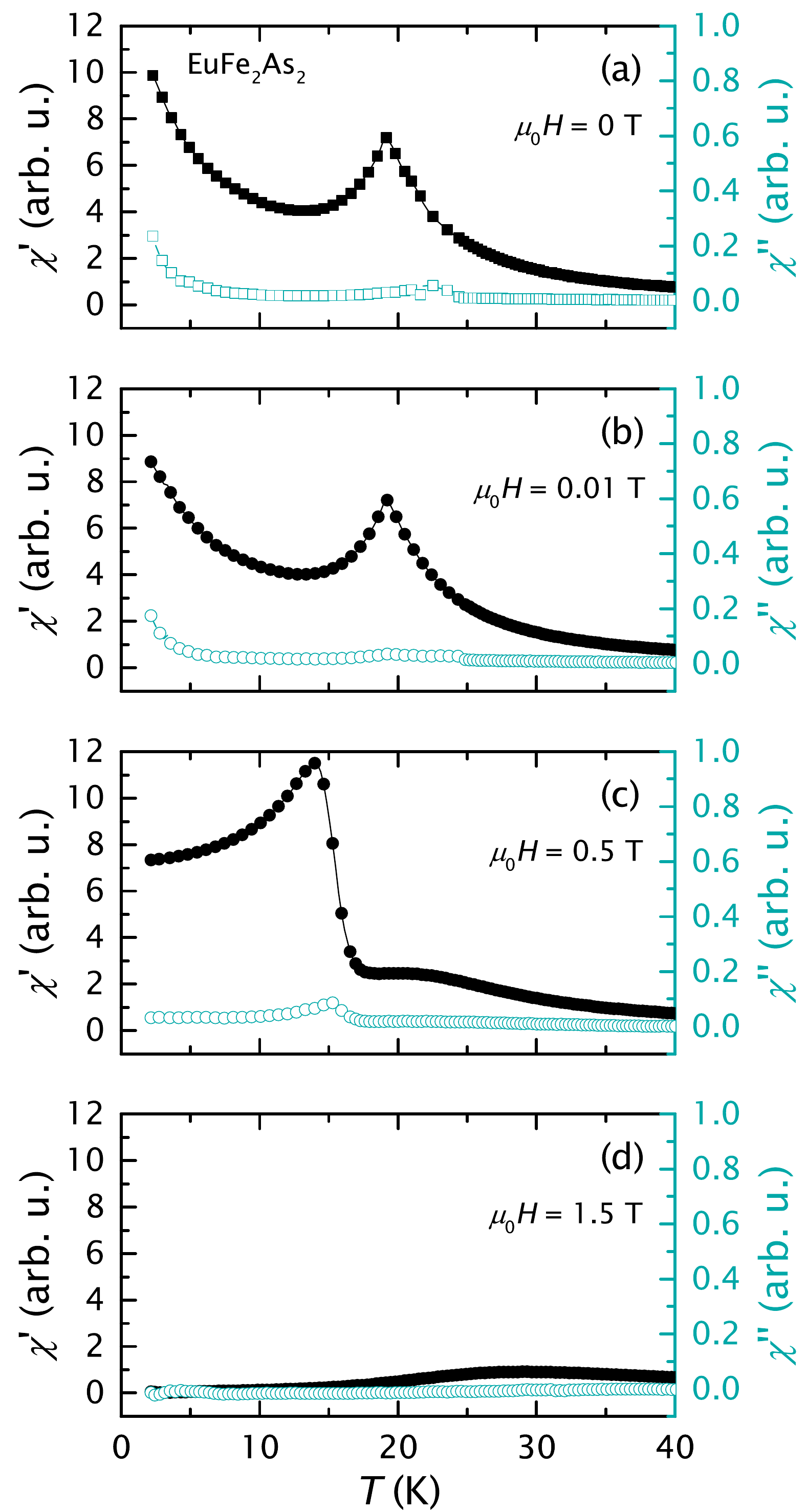}\hspace{10pt}
\includegraphics[width=0.32\linewidth]{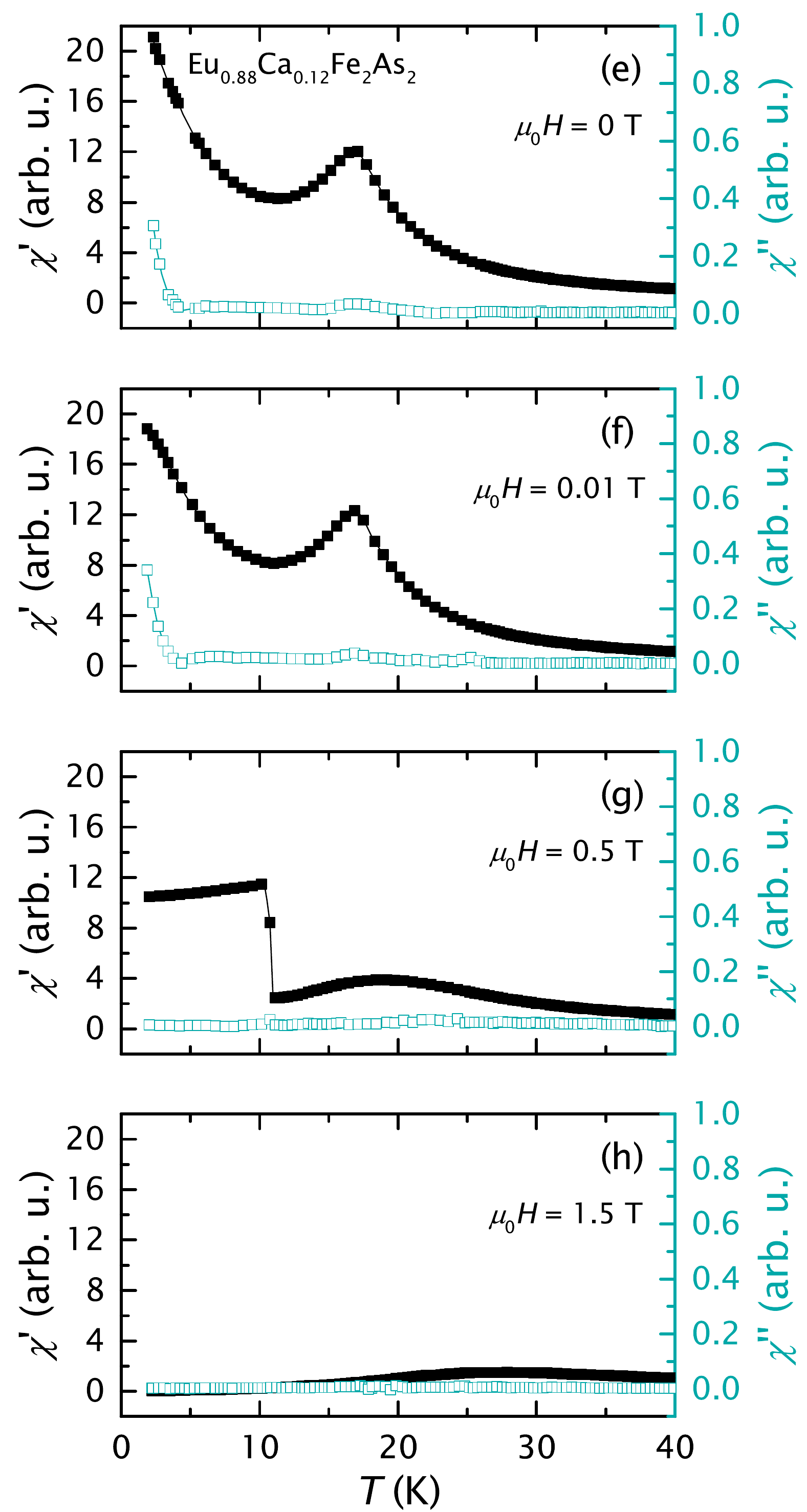}\hspace{10pt}
\includegraphics[width=0.32\textwidth]{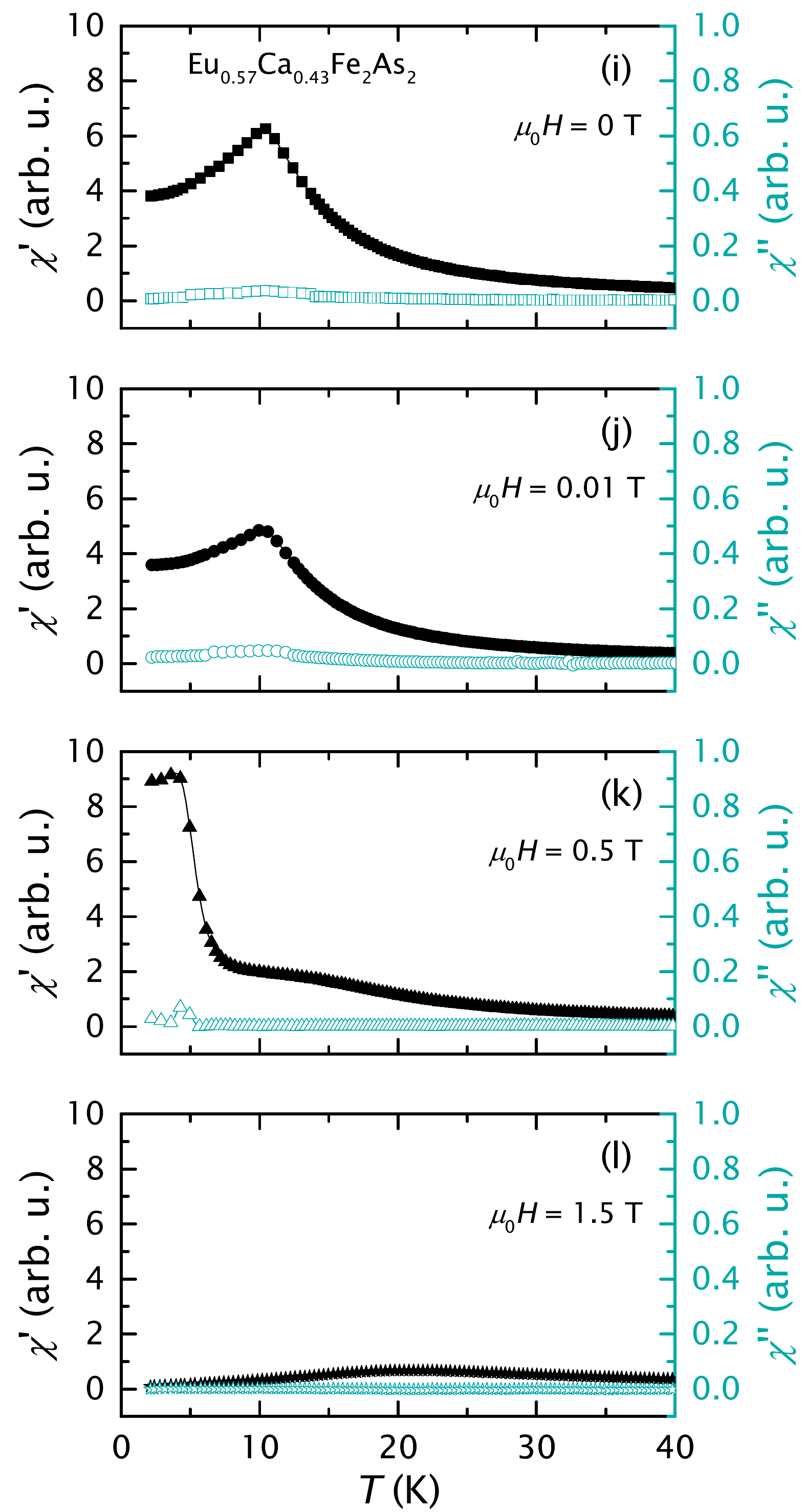}\hspace{-10pt}
\caption{Temperature dependences of the real (black) and imaginary (green) parts of ac susceptibility of (a-d) \Eu{}, (e-h) \EuCa{} and (i-l) \EuCaCa{} investigated using $\muO\Hac=\SI{10}{\milli\tesla}$ driving field with \SI{1.111}{\kilo\hertz} frequency in 0, 0.01, 0.5 and \SI{1.5}{\tesla} external magnetic fields applied perpendicular to the $\cax$\label{fig:ac-Field}}
\end{figure*}

Temperature dependences of ac susceptibility studied in 0, 0.01, 0.5 and \SI{1.5}{\tesla} external magnetic fields applied perpendicular to the $\cax$ of investigated compounds are presented on Fig.~\ref{fig:ac-Field}.

%
%


%